\documentclass[aps,pre,twocolumn,nofootinbib,tightenlines,superscriptaddress,nopacs,amsmath,amssymb,final,letterpaper]{revtex4-1}

\usepackage[utf8]{inputenc}
\usepackage{calc}
\usepackage{graphicx}
\usepackage{amsmath,amssymb,amsthm}
\usepackage{bm}
\usepackage{color}
\usepackage[dvipsnames]{xcolor}
\usepackage{enumitem}
\usepackage{tikz}
\usepackage{float}
\usepackage[percent]{overpic}
\usepackage[ruled,vlined]{algorithm2e} 

\DeclareMathOperator*{\argmax}{arg\,max}
\DeclareMathOperator*{\argmin}{arg\,min}

\newcommand{\pder}[2]{\frac{\partial #1}{\partial #2}}

\newcommand{\abs}[1]{\vert #1\vert}

\newcommand{\expec}[1]{\langle #1\rangle}

\def\multiset#1#2{\ensuremath{\left(\kern-.3em\left(\genfrac{}{}{0pt}{}{#1}{#2}\right)\kern-.3em\right)}}
\newcommand{\pluseq}{\mathrel{+}=}

\usepackage[colorlinks=true,linkcolor=blue,urlcolor=blue,citecolor=blue,anchorcolor=blue]{hyperref}

\begin{document}
\title{Fast nonparametric inference of network backbones for weighted graph sparsification}

\author{Alec \surname{Kirkley}}
\email{alec.w.kirkley@gmail.com}
\affiliation{Institute of Data Science, University of Hong Kong, Hong Kong}
\affiliation{Department of Urban Planning and Design, University of Hong Kong, Hong Kong}
\affiliation{Urban Systems Institute, University of Hong Kong, Hong Kong}

\begin{abstract}
Network backbones provide useful sparse representations of weighted networks by keeping only their most important links, permitting a range of computational speedups and simplifying network visualizations. A key limitation of existing network backboning methods is that they either require the specification of a free parameter (e.g. significance level) that determines the number of edges to keep in the backbone, or impose specific restrictions on the topology of the backbone (e.g. that it is a spanning tree). Here we develop a completely nonparametric framework for inferring the backbone of a weighted network that overcomes these limitations and automatically selects the optimal set of edges to retain using the Minimum Description Length (MDL) principle. We develop objective functions for global and local network backboning which evaluate the importance of an edge in the context of the whole network and individual node neighborhoods respectively and are generalizable to any weight distribution under Bayesian model specifications that fix the average edge weight either exactly or in expectation. We then construct an efficient and provably optimal greedy algorithm to identify the backbone minimizing our objectives, whose runtime complexity is log-linear in the number of edges. We demonstrate our methods by comparing them with existing methods in a range of tasks on real and synthetic networks, finding that both the global and local backboning methods can preserve network connectivity, weight heterogeneity, and spreading dynamics while removing a substantial fraction of edges.     
\end{abstract}


\maketitle

\section{Introduction}
In a range of practical applications including the simulation of epidemics or information cascades \cite{mercier2022effective,mathioudakis2011sparsification} and the visualization of large networks \cite{imre2020spectrum}, it is helpful to alleviate computational burden by reducing the number of edges in a network while preserving key properties of interest for computations. This task is known as \emph{graph sparsification} \cite{spielman2008graph}, and often involves identifying a \emph{network backbone} consisting of a network subgraph in which important characteristics such as degree and strength distributions \cite{serrano2009extracting} or shortest paths \cite{grady2012robust} are preserved. As there are many possible objectives for sparsification, there have been many proposed methods for identifying backbones in networks \cite{neal2022backbone,yassin2023evaluation}. Some of these methods can be used to sparsify unweighted graphs, but they most often applied to weighted graphs in which edges with higher weight are more likely to be kept in the network backbone when the weight signals edge importance with respect to the network structure or dynamics. 

A number of existing methods for graph sparsification are stochastic, aiming to preserve properties such as network spectra with high probability when sampling edges based on their structural properties \cite{spielman2011spectral,su2024generic,john2017single}. However, most network backboning methods are deterministic in nature and prune the network down to a single subgraph for analyses. Network backboning methods can generally be categorized as either \emph{global} or \emph{local} in nature \cite{yassin2023evaluation}. Global backboning methods consider the whole structure of the network when considering whether or not to remove an edge, while local backboning methods only consider the neighborhood surrounding the node(s) at one or both ends of the edge in question.

The simplest method for weighted network backboning is a global thresholding procedure in which all edges with weight less than some pre-defined threshold are dropped from the network. Another straightforward global backboning method is to compute the minimum (maximum) spanning tree with respect to weights transformed to represent some notion of distance (similarity) between nodes \cite{wu2006transport}. Spanning trees have many useful applications but are naturally acyclic so cannot possibly preserve important network properties such as transitivity, clustering, or path lengths. The High Salience Skeleton (HSS) introduced in \cite{grady2012robust} is a global backboning method that aims to identify edges critical for routing along shortest paths in the network. This method has the appealing properties of being principled and relatively insensitive to its only user-chosen threshold, but has a computational burden that is at least quadratic in the number of nodes in practice and often does not preserve the global connectivity of the graph (as we will see in Sec.~\ref{sec:synthetic}). Percolation thresholding, which imposes the maximum global weight threshold that allows the network to be a single connected component \cite{li2007transport,haimovici2013brain}, provides an alternative global backboning method which alleviates this issue at the cost of often requiring a large fraction of edges to be retained in order to ensure all nodes are connected. Other global backboning methods tend to suffer from the same issues of scalability and/or placing tight restrictions on the backbone topology, with scalability a particular issue for methods that require the computation of shortest paths \cite{zhang2018extracting,simas2021distance} or community structure \cite{rajeh2022modularity}.

Local backboning methods provide a scalable alternative to global methods by considering edges based only on nearby network structure, with the tradeoff that they may fail to preserve large-scale properties. One of the most popular methods for local network backboning is the Disparity Filter \cite{serrano2009extracting}, which drops edges if their weight falls below a pre-specified significance level under a suitable null model relative to the other edges in a node's neighborhood. The Polya Filter \cite{marcaccioli2019polya} generalizes the disparity filter to a whole family of Polya urn null models, allowing for greater flexibility in modeling weight heterogeneity. A number of other statistical methods have also been considered that aim to remove edges that are deemed insignificant under a suitable null model \cite{tumminello2011statistically,dianati2016unwinding,casiraghi2017relational,foti2011nonparametric}. All of these methods have the requirement of a user-defined threshold or significance level that directly determines the number of edges present in the final backbone. While this approach can be advantageous for identifying multiple scales at which a network exhibits significant structural patterns \cite{serrano2009extracting}, it also leaves arguably the most critical backboning choice (the number of edges to retain) up to the user and does not have a natural mechanism for selecting the single backbone representation to be used in an application. 

Properly regularized Bayesian methods for \emph{network reconstruction} have circumvented the need for fixing free parameters when inferring statistically significant network structure \cite{newman2018network,peel2022statistical,peixoto2025network}. These methods are statistically principled and robust, and are in part aimed at addressing concerns that backboning and thresholding methods---by enforcing the removal of edges---may impose unwanted biases on the structure of the pruned networks and remove important connectivity information \cite{cantwell2020thresholding}. Despite their range of benefits, network reconstruction methods are usually not suitable for network backboning because they often \emph{increase} the number of edges in a network when there is strong evidence for edges being omitted during the measurement process. They also tend to be computationally expensive due to the need to search over edge configurations involving node pairs that do not have any observed edge, the number of which scales quadratically with the network size. There are fast approximations aimed to alleviate this issue and provide better asymptotic runtime scaling \cite{peixoto2024scalable}, although the resulting scaling is still worse than that of many backboning methods due to the inherently higher complexity of the reconstruction problem. Network backboning methods' comparatively low complexity and prioritization of sparsity in the final network representation are often advantageous compared to network reconstruction techniques when considering visualization and downstream tasks with high computational overhead \cite{yassin2023evaluation}.

There are also important classes of methods aimed at sparsifying networks so as to preserve or destroy one particular structural or dynamical property. For example, spectral sparsification \cite{spielman2008graph} aims to sparsify a network while preserving the spectral properties of a particular graph operator such as the graph laplacian or graph adjacency matrix. And network dismantling \cite{braunstein2016network,zdeborova2016fast} aims to identify the minimal set of nodes or edges to remove in order to destroy any extensive connected components. While these methods are more effective than general network backbones for graph sparsification in specific problem contexts, network backbones are useful representations when a variety of exploratory and simulation-based analyses are performed on the same network, and the goal is to preserve connectivity, weight heterogeneity, and other structural and dynamical properties while reducing the dataset size for simpler manipulation and computation. As such, evaluation of network backbones is intrinsically multifaceted \cite{yassin2023evaluation}.

Here we develop fast, completely parameter-free methods for inferring global and local backbones of a weighted network based on the Minimum Description Length (MDL) principle \cite{rissanen1978modeling}. We adapt our framework to Bayesian model specifications in which total edge weight is fixed in expectation (``canonical'' models, informally) and exactly (a ``microcanonical'' model), proving that simple greedy approaches can find the exact optimum in a subset of these models. Minimizing over these objectives provides the backbone that best compresses the edge weight structure in the network, while automatically selecting for the number of edges to retain in the backbone. Our methods provide flexible, principled, and interpretable solutions for network backboning that easily scale to large networks. We test our methods on a range of real and synthetic network data, finding that they compare favorably against existing methods in a number of dimensions while eliminating the need for tuning any parameters. Finally, we study an example application in network epidemic spreading, finding that our methods allow for the accurate estimation of the associated percolation transition on networks while substantially reducing the required runtime for the calculation.

\section{Methods}
\label{sec:methods}

We start by deriving a general global backboning objective through maximum a posteriori (MAP) estimation in a Bayesian generative model of the network weights in which the total network weight is only fixed in expectation. We informally call this model a \emph{canonical} model to emphasize that the exact total weight is not known ahead of time. (A truly ``canonical'' model is more restrictive as it requires a \emph{maximum entropy} distribution with weight fixed in expectation, which in the present case is only true for the geometric and exponential distributions for integer-valued and real-valued weights respectively.) We apply this generative model to the level of node neighborhoods to develop a local backboning model. 

We then discuss the approximate equivalence between our canonical Bayesian generative model in the case of geometric weights and a simpler \emph{microcanonical} Bayesian formulation that is fixes the total network weight to an observed value. (Here we only focus on a truly ``microcanonical'' model that is uniform over all positive integer edge weight configurations with fixed total sums in the backbone and non-backbone.) Finally, we demonstrate that our microcanonical objective permits a fast, exact greedy procedure for identifying the MDL-optimal backbones, and extend our proof of greedy optimality to a set of common weight distributions under the canonical model formulation.

Our methods are motivated by the Minimum Description Length (MDL) principle, which states that the optimal model for a dataset is the model that allows us to transmit the dataset using the fewest bits \cite{rissanen1978modeling}. By considering both the information to transmit a dataset given a model \emph{and} the information to transmit the model itself, the MDL principle provides a flexible framework for penalizing overly complex models to reduce the risk of overfitting. The MDL principle has been employed in the network context for a number of methods in the past, for the tasks of community detection \cite{peixoto2019bayesian,kirkley2022spatial,peixoto2023implicit}, graph comparison \cite{coupette2021graph,coupette2022differentially}, graph summarization \cite{liu2018graph,kirkley2023compressing}, and for extracting structural regularities in dynamic networks and hypergraphs \cite{shah2015timecrunch,young2021hypergraph,kirkley2024identifying,kirkley2024inference}. In the backboning context, the MDL principle allows us to naturally promote sparsity in the inferred backbone while encouraging the backbone to carry the most significant weights in the network. There are many possible weight distributions for our canonical backboning models, each resulting in a different MDL formulation, and so technically we are considering not one but \emph{many} possible MDL formulations of the backboning problem.

\subsection{Canonical Models for Global and Local Backboning}
\label{sec:canonical}

Our problem considers an input network $G=\{(i,j,w_{ij})\}$, which we represent as a set of $E$ weighted tuples $e=(i,j,w_{ij})$. The tuple $e=(i,j,w_{ij})$ indicates an edge from a node $i$ to a node $j$ with weight $w_{ij} \geq 0$ that represents the strength of interaction among $i$ and $j$. The network $G$ must be weighted, as the backboning methods we present find backbones that allow for compression of the edge weights in $G$. The graph $G$ may have self-edges or multi-edges---multi-edges can be combined into a single edge weight---and $G$ may be directed or undirected. In the undirected case we simply consider both $(i,j,w_{ij})$ and $(j,i,w_{ij})$ to be in $G$, with the edge $(i,j,w_{ij})$ appearing in an undirected backbone if either $(i,j,w_{ij})$ or $(j,i,w_{ij})$ is retained in the directed backbone (as done in \cite{serrano2009extracting}). Bipartite networks can be accommodated by our framework with no additional modifications. We will let $W=\sum_{e\in G}w_{e}$ denote the total edge weight in the weighted graph $G$. The goal of our network backboning method is to extract a subset $G^{(b)}\subseteq G$ of weighted edges that best compress the overall weight distribution in $G$. We call the subset $G^{(b)}$ the \emph{backbone} of $G$, which has $E^{(b)}\leq E$ edges with a total weight of $W^{(b)}\leq W$. 

To derive a global backboning objective, we can consider a generative process in which we start with an unweighted network $G_u$, and for each edge $e\in G_u$ we classify $e$ as being either a backbone edge or non-backbone edge by assigning it an indicator variable
\begin{align}
b_e=\begin{cases}
1,~~e\text{ is a backbone edge},\\
0,~~e\text{ is not a backbone edge}.
\end{cases}    
\end{align}
This results in a binary vector $\bm{b}$ indexed by edge pairs $e=1,...,E$ that partitions the set of edges into a backbone and non-backbone. We then use a mixture model for the edge weights in which the weight $w_{e}$ is generated according to the distribution $P(w_{e}\vert \bm{\theta}_{b_{e}})$, where $\bm{\theta}_{b_{e}}$ is a set of parameters for the distribution $P$ that depend on whether or not the edge $e$ is in the backbone. We therefore have two independent sets of model parameters $\bm{\theta}_1$ and $\bm{\theta}_0$ for the backbone and non-backbone edges respectively. This mixture model imposes a distinction between the weight statistics of the backbone and the non-backbone edges, which is the essential feature utilized by the majority of weighted network backboning methods \cite{serrano2009extracting,tumminello2011statistically,dianati2016unwinding,casiraghi2017relational,foti2011nonparametric,marcaccioli2019polya,li2007transport}.

For the canonical model, we generate the backbone assignments $\bm{b}\sim \text{Bernoulli}(\pi_b)$, where $\pi_b\in [0,1]$ is the probability of an edge being assigned to the backbone. This gives us a prior distribution over the backbone $\bm{b}$ of
\begin{align}
P(\bm{b}\vert \pi_b) = \prod_{e\in G_u}\pi_b^{b_{e}}(1-\pi_b)^{1-b_{e}}.    
\end{align}
The likelihood of an observed weighted network $G$ and a backbone assignment $\bm{b}$ given the model parameters and the unweighted network $G_u$ can then be written as
\begin{align}\label{eq:likelihood}
P(G,\bm{b}\vert &\bm{\theta}_1,\bm{\theta}_0,\pi_b) 
=P(G\vert \bm{\theta}_1,\bm{\theta}_0,\bm{b})P(\bm{b}\vert \pi_b)\nonumber\\
&= \prod_{e\in G_u}[\pi_bP(w_{e}\vert \bm{\theta}_1)]^{b_{e}}[(1-\pi_b)P(w_{e}\vert \bm{\theta}_0)]^{1-b_{e}}\nonumber \\
&= \pi_b^{E^{(b)}}(1-\pi_b)^{E-E^{(b)}}\\
&\times \prod_{e\in G^{(b)}(\bm{b})}P(w_{e}\vert \bm{\theta}_1)\prod_{e\in G\setminus G^{(b)}(\bm{b})}P(w_{e}\vert \bm{\theta}_0), \nonumber
\end{align} 
where $G^{(b)}(\bm{b})\subseteq G$ is the subset of weighted backbone edges associated with the assignment $\bm{b}$. The above likelihood is (as implied in the model formulation) conditional on knowing the unweighted network $G_u$, but we remove explicit dependence on $G_u$ in the notation for brevity.

Putting priors $P(\bm{\theta}_1)$, $P(\bm{\theta}_0)$, and $P(\pi_b)$ on the model parameters that are conjugate to the likelihoods $\prod_{e}P(w_{e}\vert \bm{\theta}_1)$, $\prod_{e}P(w_{e}\vert \bm{\theta}_0)$, and $P(\bm{b}\vert \pi_b)$ respectively, we can marginalize over $\{\bm{\theta}_1,\bm{\theta}_0,\pi_b\}$ to form the joint distribution for the canonical model
\begin{align}\label{eq:jointcanonicalglobal}
P^{(\text{global})}_{C}(G,\!\bm{b}) &=\!\! \int\!\!  P(G,\bm{b}\vert \bm{\theta}_1,\bm{\theta}_0,\pi_b)P(\bm{\theta}_1)P(\bm{\theta}_0)P(\pi_b)d\bm{\theta} d\pi_b   \nonumber \\
&= B(E^{(b)}+1,E-E^{(b)}+1) \nonumber\\
&\times \int P(\bm{\theta}_1) \prod_{e\in G^{(b)}}P(w_{e}\vert \bm{\theta}_1)d\bm{\theta}_1\\
&\times\int P(\bm{\theta}_0) \prod_{e\in G\setminus G^{(b)}} P(w_{e}\vert \bm{\theta}_0)d\bm{\theta}_0, \nonumber
\end{align}
where 
\begin{align}
B(x,y) = \frac{(x-1)!(y-1)!}{(x+y-1)!}  
\end{align}
is the Beta function.

Maximum a posteriori (MAP) estimation with this model then aims to solve the following optimization problem to identify the optimal backbone assignments $\bm{\hat b}$
\begin{align}\label{eq:MAP}
\bm{\hat b}
&= \argmax_{\bm{ b}\in \{0,1\}^{E}} \left\{P(\bm{b}\vert G)\right\}.
\end{align}
This can be equivalently framed in terms of the \emph{description length} $\mathcal{L}^{(\text{global})}_{C}(G^{(b)})$ of the canonical global backbone model, which we can parametrize with just the weighted backbone $G^{(b)}$ as
\begin{align}\label{eq:DLglobalfromBayes}
\mathcal{L}^{(\text{global})}_{C}(G^{(b)}) &= -\log P^{(\text{global})}_{C}(G,\bm{b}) \nonumber \\
&= \log(E+1) + \log{E\choose E^{(b)}} \nonumber\\
&- \log \left[\int P(\bm{\theta}_1) \prod_{e\in G^{(b)}}P(w_{e}\vert \bm{\theta}_1)d\bm{\theta}_1\right]\\
&-\log  \left[\int P(\bm{\theta}_0) \prod_{e\in G\setminus G^{(b)}} P(w_{e}\vert \bm{\theta}_0)d\bm{\theta}_0\right].\nonumber
\end{align}
We use the notation $\log(x) \equiv \log_2(x)$ for brevity, so that the description length is in units of bits. Eq.~\ref{eq:DLglobalfromBayes} tells us, for an optimal encoding over networks and backbones under our assumptions about the weight distribution, how much information is required to transmit a specific network $G$ and backbone partition $\bm{b}$ \cite{mackay2003information}. 

Solving Eq.~\ref{eq:MAP} is equivalent to minimizing the description length of Eq.~\ref{eq:DLglobalfromBayes}, giving the minimum description length (MDL) objective
\begin{align}
\hat G^{(b)} = \argmin_{G^{(b)}\subseteq G} \left\{\mathcal{L}_{C}^{(\text{global})}(G^{(b)}) \right\},    
\end{align}
where $\hat G^{(b)}$ is the MDL-optimal backbone. We can see that Eq.~\ref{eq:DLglobalfromBayes} has two contributions. The first two terms penalize the complexity of the backbone $G^{(b)}$ by encouraging the backbone size $E^{(b)}$ to be small, for $E^{(b)}\leq E/2$. These terms also encourage the backbone sizes $E^{(b)}$ to be large, for $E^{(b)}> E/2$---the backbone $G^{(b)}$ is equally as compressive as its complement $G\setminus G^{(b)}$ since specifying the backbone edges is equivalent to specifying the non-backbone edges given that the unweighted graph $G_u$ is known. (This ``bit flip symmetry'' under the interchange $0\Leftrightarrow 1$ is intrinsic to information theoretic methods such as mutual information measures \cite{felippe2024network}.) However, as the backbones $G^{(b)}$ with sizes in the regime $E^{(b)}\geq E/2$ are nearly as dense as the original graphs, they do not provide meaningful sparsification, so we will focus on the regime $E^{(b)}\leq E/2$---it is arguably not meaningful to suggest that more than half of the edges in $G$ have statistically significant weights, regardless of the null model one chooses. The final two terms in Eq.~\ref{eq:DLglobalfromBayes} encourage the weight distributions in the backbone and non-backbone to be distinct and for the model $P$ to provide an effective summary of their variabilities.

We can generalize our canonical Bayesian model to the neighborhood level for the edge weights in each node neighborhood $\partial_i=\{(i,j,w_{ij})\}$ for all nodes $j$ towards which $i$ has a directed or undirected edge. We denote the strength $s_i$ of each node $i$ as its total incident weight $s_i=\sum_{e\in \partial_i}w_{e}$ and the degree of node $i$ with $k_i=\abs{\partial_i}$. In this model, each node $i$ has its own probability $\pi_b(i)$ that one of its adjacent edges $e\in \partial_i$ is in the backbone $G^{(b)}$. Similarly, each node $i$ has its own model parameters $\bm{\theta}_1(i)$ and $\bm{\theta}_0(i)$ specific to its neighborhood's edge weights. This parametrization allows the network neighborhoods $\partial_i$ to have different weight distributions, which in turn allows the local backboning procedure to identify statistically significant weights in each neighborhood separately as is done in other local backboning methods \cite{serrano2009extracting}.

We can then compute the joint distribution $P^{(\text{local})}_{C}(G,\bm{b})$ for this canonical local backboning model by integrating out the parameters $\{\pi_b(i),\bm{\theta}_1(i),\bm{\theta}_0(i)\}$ for each node neighborhood $\partial_i$ and taking a product over the independent neighborhoods, assuming a directed network where $(i,j)$ and $(j,i)$ are treated independently. The result is equivalent to assuming $N$ independent copies of the global backboning method, one for each neighborhood $\partial_i$, which each have their own associated backbone $\partial_i^{(b)}$ with total weight $s_i^{(b)}$ and number of edges $\abs{\partial_i^{(b)}}=k_i^{(b)}$. The resulting joint distribution is given by
\begin{align}\label{eq:jointcanonicallocal}
P^{(\text{local})}_{C}(G,\bm{b}) &= \prod_{i=1}^{N}P_C^{(\text{global})}(\partial_i,\bm{b}_{\partial_i}) \nonumber\\
&= \prod_{i=1}^{N}B(k_i^{(b)}+1,k_i-k_i^{(b)}+1) \\
&\times \prod_{i=1}^{N}\int P(\bm{\theta}_1(i)) \prod_{e\in \partial^{(b)}_i}P(w_{e}\vert \bm{\theta}_1(i))d\bm{\theta}_1(i) \nonumber\\
&\times \prod_{i=1}^{N}\int P(\bm{\theta}_0(i))\!\!\prod_{e\in \partial_i\setminus \partial^{(b)}_i} \!\! P(w_{e}\vert \bm{\theta}_0(i))d\bm{\theta}_0(i), \nonumber
\end{align}
where $\bm{b}_{\partial_i}=\{b_e:e\in \partial_i\}$ is the backbone partition of the edges in the neighborhood $\partial_i$. The description length of interest is then
\begin{align}\label{eq:DLlocalfromBayes}
\mathcal{L}^{(\text{local})}_C(G^{(b)}) = -\log P^{(\text{local})}_{C}(G,\bm{b}) = \sum_{i=1}^{N} \mathcal{L}_C^{(\text{neig})}(\partial_i^{(b)}),   
\end{align}
where 
\begin{align}\label{eq:neigdlcanonical}
\mathcal{L}^{(\text{neig})}_C&(\partial_i^{(b)}) = \log(k_i+1) + \log {k_i\choose k_i^{(b)}} \nonumber\\
&~~- \log \left[\int P(\bm{\theta}_1(i)) \prod_{e\in \partial^{(b)}_i}P(w_{e}\vert \bm{\theta}_1(i))d\bm{\theta}_1(i)\right]\\
&~~- \log \left[\int P(\bm{\theta}_0(i)) \!\!\prod_{e\in \partial_i\setminus \partial^{(b)}_i}\!\!P(w_{e}\vert \bm{\theta}_0(i))d\bm{\theta}_0(i)\right]\nonumber.
\end{align}
Similarly to the global case, the local objective has bit flip symmetry, but in this case at the neighborhood level. We therefore focus on the regime $k_i^{(b)}\leq k_i/2$, to promote sparse backbones---under the same logic as before, one cannot say that more than half of the edges in a neighborhood have statistically significant weights relative to the rest of the neighborhood, under any reasonable null model that treats nodes independently. 

Eq.~\ref{eq:DLglobalfromBayes} and Eq.~\ref{eq:DLlocalfromBayes} provide flexible objectives to solve for the backbone $G^{(b)}$ of a weighted network $G$ by assessing global and local edge weight heterogeneity respectively. These canonical model objectives allow for any underlying assumption about the weight distributions within and outside of the backbone $G^{(b)}$. They are adaptable to either discrete or continuous weights, and provide a means of model selection among different mixtures of underlying weights through the comparison of the description length values under different distributional assumptions for $P$.

\subsection{Microcanonical Models for Global and Local Backboning}
\label{sec:microcanonical}

An alternative Bayesian formulation of network backbones can be done in a \emph{microcanonical} framework in which we fix the network weight $W$ rather than observing $W$ as a result of generating the network edge weights. This allows us to calibrate the model so that the weight generative process being fit to the observed data $G$ will always generate the correct total weight $W$. This has the benefit of not requiring the marginalization over any free parameters. Additionally, fixing statistics of interest exactly rather than in expectation (and vice versa) can permit easier analytical treatment as well as result in qualitatively distinct statistical behaviors in network ensembles \cite{voitalov2020weighted}, making microcanonical models worth exploring in their own right.

For the microcanonical generative model of network backbones, we assume that $W$ is known in addition to the unweighted graph $G_u$. We first draw the number of backbone edges $E^{(b)}$ uniformly at random from the interval $[0,E]$, giving a prior of $P(E^{(b)})=(E+1)^{-1}$. Similarly, we draw the total backbone weight $W^{(b)}$ uniformly at random. Given that we have positive integer weights, we know that $W^{(b)}\geq E^{(b)}$, and $W^{(b)}\leq W-(E-E^{(b)})$, since each edge must have weight at least $1$. Therefore, given $E^{(b)}$, $W^{(b)}$ is then drawn uniformly at random from the interval $[E^{(b)},W-(E-E^{(b)})]$, resulting in the prior $P(W^{(b)}\vert E^{(b)})=(W-E+1)^{-1}$. We then draw the backbone assignments $\bm{b}$ uniformly at random from the set of all binary partitions with $\sum_{e=1}^{E}b_e=E^{(b)}$, of which there are ${E\choose E^{(b)}}$, giving a prior of $P(\bm{b}\vert E^{(b)})={E\choose E^{(b)}}^{-1}$. Finally, given the specification of the backbone edges and total backbone weight, we draw the weights of all edges in the backbone and non-backbone uniformly at random given that the $E^{(b)}$ edges in the backbone must have weights that sum to $W^{(b)}$ and the $E-E^{(b)}$ edges in the non-backbone must have weights that sum to $W-W^{(b)}$. The number of configurations of $k$ positive integers that sum to $n$ is called the number of \emph{compositions} of the integer $n$ into $k$ parts, and is given by ${n-1\choose k-1}$, so the probability of a particular configuration of the backbone and non-backbone weights given their sums is $P(G\vert 
 \bm{b},W^{(b)},E^{(b)})={W^{(b)}-1\choose E^{(b)}-1}^{-1}{W-W^{(b)}-1\choose E-E^{(b)}-1}^{-1}$.

Putting this all together, the full joint distribution of the microcanonical model can be written as
\begin{align}\label{eq:jointmicrocanonicalglobal}
P^{(\text{global})}_{M}(G,\!\bm{b})
\!&=\! P(E^{(b)})\!\times\! P(W^{(b)}\vert E^{(b)})\!\times\! P(\bm{b}\vert E^{(b)})\nonumber\\
 &~~~~~\!\times\! P(G\vert \bm{b},W^{(b)},E^{(b)}) \\
 &\!=\!(E\!+\!1)^{-1}\!\times\! (W\!-\!E\!+\!1)^{-1}\!\times\! {E\choose E^{(b)}}^{-1} \nonumber\\
 &~~~~~\!\times\! {W^{(b)}\!-\!1\choose E^{(b)}\!-\!1}^{-1}{W\!-\!W^{(b)}\!-\!1\choose E\!-\!E^{(b)}\!-\!1}^{-1}\nonumber,   
\end{align}
which has description length
\begin{align}\label{eq:DLglobalmicro}
\mathcal{L}_M^{(\text{global})}\!(G^{(b)})
\!&=\!-\!\log P^{(\text{global})}_{M}(G,\bm{b}) \\
&\!=\!\log (E+1) \!+\! \log (W\!-\!E\!+\!1) \!+\! \log {E\choose E^{(b)}} \nonumber\\
 &~+\log {W^{(b)}\!-\!1\choose E^{(b)}\!-\!1} \!+\!\log{W-W^{(b)}\!-\!1\choose E\!-\!E^{(b)}-1}\nonumber.
\end{align}
As before, this objective can be minimized over backbones $G^{(b)}$ to find the MDL-optimal backbone under the microcanonical global model.

We can also formulate a microcanonical local backbone model in a similar way, drawing $\{k_i^{(b)},s_i^{(b)},\bm{b}_{\partial_i},\partial^{(b)}_i,\partial_i\setminus \partial^{(b)}_i\}$ using hierarchical uniform priors analogous to those for $\{E^{(b)},W^{(b)},\bm{b},G^{(b)},G\setminus G^{(b)}\}$ respectively, to form each neighborhood-level term $P^{(\text{global})}_M(\partial_i^{(b)},\bm{b}_{\partial_i}\vert s_i)$ in an independent factorization analogous to Eq.~\ref{eq:jointcanonicallocal}. The only additional consideration we must make for the microcanonical model is that, since $s_i$ is assumed known for all $i=1,...,N$, we need to put a uniform prior on $\bm{s}=\{s_1,...,s_N\}$ given the known total network weight $W$. By adding this prior, we can compare the global and local microcanonical description lengths directly, since both models assume the same prior knowledge ($G_u$ and $W$). Since each edge has weight at least $1$, node $i$ must have a strength that satisfies $s_i\geq k_i$. In other words, a uniform distribution of strengths $\bm{s}$ will consider all possible ways to distribute the \emph{excess weight} $W-E$ onto the $N$ nodes, allowing for some nodes to receive no excess weight (in other words, $s_i=k_i$). A uniform prior on $\bm{s}$ is then equivalent to generating a random composition of $W-E$ as $N$ non-negative integers, allowing the integers to potentially be zero, of which there are ${N+W-E-1\choose W-E}$ possible combinations. The final prior for $\bm{s}$ is then $P(\bm{s})={N+W-E-1\choose W-E}^{-1}$. This prior is a global constant that does not impact the inferred backbones, but is technically necessary for a direct comparison of the global and local description length values for model comparison. 

Including the prior on the strengths $P(\bm{s})$, we have a description length for the local microcanonical model of
\begin{align}\label{eq:DLlocalmicro}
\mathcal{L}^{(\text{local})}_M&(G^{(b)}) = \!-\!\log P(\bm{s}) \!-\!\log P^{(\text{local})}_M(G^{(b)},\bm{b}\vert \bm{s}) \\
&= \!-\!\log P(\bm{s}) \!-\!\log \left(\prod_{i=1}^{N}P^{(\text{global})}_M(\partial_i^{(b)},\!\bm{b}_{\partial_i}\vert s_i)\right) \nonumber\\
&= \log {N\!+\!W\!-\!E\!-\!1\choose W\!-\!E}+\sum_{i=1}^{N} \mathcal{L}_M^{(\text{neig})}(\partial_i^{(b)}) \nonumber,   
\end{align}
where 
\begin{align}\label{eq:neigdlmicro}
\mathcal{L}^{(\text{neig})}_M(\partial_i^{(b)}) &=\log (k_i\!+\!1) + \log (s_i\!-\!k_i\!+\!1) + \log {k_i\choose k_i^{(b)}} \nonumber\\
 &~+\log {s_i^{(b)}-1\choose k_i^{(b)}-1} +\log{s_i-s_i^{(b)}-1\choose k_i-k_i^{(b)}-1}.
\end{align}
One can show (see Appendix~\ref{appendix:equivalence}) that the microcanonical global and neighborhood-level description lengths of Eq.~\ref{eq:DLglobalmicro} and Eq.~\ref{eq:neigdlmicro} are asymptotically equivalent to the canonical description lengths of Eq.~\ref{eq:DLglobalfromBayes} and Eq.~\ref{eq:neigdlcanonical} respectively when weights follow a geometric distribution. The full canonical local description length of Eq.~\ref{eq:DLlocalfromBayes} with geometric weights is only equivalent to the full microcanonical local description length of Eq.~\ref{eq:DLlocalmicro} up to the term $\log {N+W-E-1\choose W-E}$ associated with the prior $P(\bm{s})$, which is a global constant that does not affect the inferred backbones.

One of the major benefits of using MDL-based learning objectives is that they naturally provide a simple criterion for model selection---the model providing the lowest description length at the end of the fitting process is the best model for the data \cite{rissanen1978modeling}. We can then determine whether a global or local backbone, as well as which weight model, provides the most parsimonious description of the observed network $G$, by fitting the objectives in Eq.~\ref{eq:DLglobalfromBayes}, Eq.~\ref{eq:DLlocalfromBayes}, Eq.~\ref{eq:DLglobalmicro}, and Eq.~\ref{eq:DLlocalmicro} then comparing the resulting description lengths. The common downside of MDL objectives is that they can be challenging to minimize, often requiring sophisticated Markov chain Monte Carlo methods \cite{peixoto2017nonparametric}. However, we show in the next section that the microcanonical description lengths in Eq.~\ref{eq:DLglobalmicro} and Eq.~\ref{eq:DLlocalmicro} can be minimized exactly using a fast greedy algorithm. We can extend this proof of optimality to a set of common weight distributions under the canonical formulations in Eq.~\ref{eq:DLglobalfromBayes} and Eq.~\ref{eq:DLlocalfromBayes}.

\subsection{Optimization}
\label{sec:optimization}

All four of the MDL backboning objectives of Sec.~\ref{sec:canonical} and Sec.~\ref{sec:microcanonical}---Eq.s~\ref{eq:DLglobalfromBayes},~\ref{eq:DLlocalfromBayes},~\ref{eq:DLglobalmicro}, and~\ref{eq:DLlocalmicro}---must be minimized over sets of backbone edges $G^{(b)}\subseteq G$ to identify an MDL-optimal backbone under each description length formulation. With na\"ive optimization by enumeration of all possible backbones $G^{(b)}$, this would require $2^E$ evaluations of a given objective---one for each possible backbone---which is intractable for all but the smallest networks. We show here that for the microcanonical objectives of Eq.s~\ref{eq:DLglobalmicro} and~\ref{eq:DLlocalmicro}, as well as for a wide variety of weight distributions for the canonical models of Eq.s~\ref{eq:DLglobalfromBayes} and~\ref{eq:DLlocalfromBayes}, one can find the MDL-optimal backbone using simple greedy procedures requiring only $O(E\log E)$ operations---the bottleneck consisting of sorting the set of edges $G$ by weight at the beginning of the algorithm. 

We can analyze the microcanonical objectives first, then generalize our results to the canonical objectives. Consider a greedy procedure in which the backbone $G^{(b)}$ starts empty and we add edges $e\in G$ to $G^{(b)}$ in decreasing order of weight $w_e$, stopping the procedure when $E^{(b)}=E/2$ edges have been added to the backbone. (As discussed in Sec.~\ref{sec:canonical}, given the bit flip symmetry of the description length objectives with respect to $\bm{b}$, the regime of interest for backboning is $E^{(b)}\leq E/2$.) We then scan over the past solution candidates and identify the one with the lowest description length to find the optimal backbone. Details of this procedure are shown in Algorithm~\ref{alg:global}. During this algorithm, at all steps we must necessarily have that 
\begin{align}\label{eq:ineqgreedy}
\frac{W^{(b)}}{E^{(b)}}\geq \frac{W}{E} \geq \frac{W-W^{(b)}}{E-E^{(b)}}.    
\end{align}
In other words, the average edge weight in the backbone $G^{(b)}$ is at least as high as the average edge weight in the total network $G$, which is in turn at least as high as the average edge weight in the non-backbone $G\setminus G^{(b)}$. 

We can now consider the change in description length
\begin{align}\label{eq:deltaDL}
\Delta \mathcal{L}_M&^{(\text{global})}(W^{(b)}\pluseq 1)
= \mathcal{L}_M^{(\text{global})}(G_+^{(b)})- \mathcal{L}_M^{(\text{global})}(G^{(b)}) \nonumber\\
&= \log {\frac{W^{(b)}}{W^{(b)}-E^{(b)}+1}\frac{(W-W^{(b)})-(E-E^{(b)})}{W-W^{(b)}-1}},    
\end{align}
where $E_+^{(b)}=E^{(b)}\geq 1$ and $W_+^{(b)}=W^{(b)}+1$ for the backbone $G_+^{(b)}$. Eq.~\ref{eq:deltaDL} gives the change in description length induced by an increase in the total backbone weight $W^{(b)}$ for a fixed number of backbone edges $E^{(b)}$. Rearranging the argument of the logarithm, we can see that $\Delta \mathcal{L}_M^{(\text{global})}(W^{(b)}\pluseq 1) < 0$---in other words, the description length is decreasing in $W^{(b)}$---when
\begin{align}\label{eq:ineqdecreasing}
W^{(b)} > (E^{(b)}-1)\frac{W-1}{E-2}.    
\end{align}

Now, using the inequality of Eq.~\ref{eq:ineqgreedy} and $1\leq E^{(b)}\leq E/2$, we have that for the greedy procedure in Algorithm~\ref{alg:global}
\begin{align}
W^{(b)}
&=(E^{(b)}-1)\frac{W^{(b)}}{E^{(b)}}\frac{E^{(b)}}{E^{(b)}-1} \nonumber\\
&\geq (E^{(b)}-1)\frac{W}{E}\frac{E}{E-2} \nonumber\\
&> (E^{(b)}-1)\frac{W-1}{E-2},
\end{align}
which is precisely the condition required in Eq.~\ref{eq:ineqdecreasing} for the description length $\mathcal{L}_M^{(\text{global})}$ to be deceasing in the backbone weight $W^{(b)}$ for fixed $E^{(b)}$. In other words, the best backbone $G^{(b)}$ for any number of edges $E^{(b)}\in [1,E/2]$ is the one with the highest total weight $W^{(b)}$, and so Algorithm~\ref{alg:global} will always be able to find an optimal solution for the backbone $G^{(b)}$. (The other optimal solution has backbone $G\setminus G^{(b)}$, due to the bit flip symmetry.) Note that this condition still allows for the backbone $G^{(b)}$ to potentially be empty---the description length may be minimized at $E^{(b)}=0$, if the weights are sufficiently homogeneous. Using a similar argument, we can also show that the canonical description length of Eq.~\ref{eq:DLglobalfromBayes} is minimized with the greedy procedure in Algorithm~\ref{alg:global} for common weight distributions including the Poisson, Geometric, and Exponential distributions, which all fall under the Natural Exponential Family (NEF) of distributions \cite{morris1982natural} (see Appendix~\ref{appendix:canonical-proof}). 

Mapping the global variables $\{E^{(b)},W^{(b)},\bm{b},G^{(b)},G\setminus G^{(b)}\}$ to the neighborhood level variables $\{k_i^{(b)},s_i^{(b)},\bm{b}_{\partial_i},\partial^{(b)}_i,\partial_i\setminus \partial^{(b)}_i\}$ as before, we have that the best neighborhood-level backbone $\partial_i^{(b)}$ for any node degree $k_i^{(b)}\in [1,k_i/2]$ is the one with the highest total weight $s_i^{(b)}$, and so the application of the greedy procedure in Algorithm~\ref{alg:global} to each neighborhood $\partial_i$ separately gives the MDL-optimal backbone $G^{(b)}=\cup_{i=1}^{N}\partial_i^{(b)}$ for the local objectives of Eq.~\ref{eq:DLlocalfromBayes} and Eq.~\ref{eq:DLlocalmicro}. Details of this procedure are shown in Algorithm~\ref{alg:local}.

\begin{algorithm}[t]
    \SetAlgoLined
    \KwIn{
    Graph $G=\{(e=(i,j),w_{e}=w_{ij})\}_{e=1}^{E}$
    }
    \KwOut{Global backbone $G^{(b)}\subseteq G$}
    \BlankLine
    \textbf{Initialization:}\\
    \Indp
        -Sort edges $G$ by decreasing edge weight \indent $w_e$\;
    \Indm

    \BlankLine
    \textbf{Algorithm:}\\
    \Indp
        -Set $W^{(b)}=0$, $G^{(b)}=[~]$, $\mathcal{L}s=[~]$\;
        -Compute $\mathcal{L}=\mathcal{L}^{(\text{global})}(W^{(b)}=0,E^{(b)}=0)$ using Eq.~\ref{eq:DLglobalfromBayes} or Eq.~\ref{eq:DLglobalmicro} (whichever is being minimized) and append to $\mathcal{L}s$\;     
    
    \BlankLine
    \For{$E^{(b)} \gets 1$ \KwTo $\lfloor E/2\rfloor$}{
        ~~~~~~$i,j,w_{ij} \gets G[E^{(b)}-1]$\;
        ~~~~~~$W^{(b)} \pluseq w_{ij}$\;
        ~~~~~~Append $(i,j,w_{ij})$ to $G^{(b)}$\;
        ~~~~~~$\mathcal{L} \gets \mathcal{L}^{(\text{global})}(W^{(b)},E^{(b)})$\;
        ~~~~~~Append $\mathcal{L}$ to $\mathcal{L}s$\;
    }
    \BlankLine
    \Indp
        -Set $G^{(b)} \gets G^{(b)}[:\text{argmin}(\mathcal{L}s)] $\;
                
    \Indm

    \BlankLine
    
    \Return{$G^{(b)}$}\;
    \caption{Global backboning (Eq.~\ref{eq:DLglobalfromBayes},  Eq.~\ref{eq:DLglobalmicro})}
    \label{alg:global}
\end{algorithm}

\begin{algorithm}[t]
      \SetAlgoLined
    \KwIn{
    -Graph $G=\{(e=(i,j),w_{e}=w_{ij})\}_{e=1}^{E}$\;
    ~~~~~~~~~~~~-directed (boolean)
    }
    \KwOut{Local backbone $G^{(b)}\subseteq G$}
    \BlankLine
    \textbf{Initialization:}\\
    \Indp
        ~~~-If directed is False, duplicate edges in $G$ to include both directions $(i,j)$ and $(j,i)$\;
        ~~~-Sort edges $G$ by decreasing edge weight \indent $w_e$\;
        ~~~-Iterate through $G$ and construct adjacency list $a$ such that $a[i]=\partial_i$. $a[i]$ is sorted by weight since $G$ was sorted\;   
    \Indm
    \BlankLine
    \textbf{Algorithm:}\\
    -Set $G^{(b)}=[~]$\;
    \For{$i\gets 1$ \KwTo $N$}{
        ~~~~~~Compute $\partial_i^{(b)}$ by applying Algorithm 1 to $a[i]$\;
        ~~~~~~~~~~-Ignore initialization step\;
        ~~~~~~~~~~-Use associated global model ($C/M$) for $\mathcal{L}$\;
        ~~~~~~Append $\partial_i^{(b)}$ to $G^{(b)}$\;
    }
    \BlankLine    
    ~~~-If directed is False, sort node indices in each edge $(i,j,w_{ij})\in G^{(b)}$ and remove duplicates to convert back to undirected graph\;
    \BlankLine
    \Return{$G^{(b)}$}\;
    \caption{Local backboning (Eq.~\ref{eq:DLlocalfromBayes},  Eq.~\ref{eq:DLlocalmicro})}
    \label{alg:local}
\end{algorithm}

The computational complexity of both Algorithm~\ref{alg:global} and Algorithm~\ref{alg:local} is $O(E\log E)$, with the bottleneck being the sorting of $G$ during initialization. Both algorithms are therefore only slightly slower in practice than constructing $G$ in the first place. These algorithms are also completely nonparametric, with no required threshold for determining the edges to retain in the backbone, as the optimal edge set is determined automatically through parsimony and the MDL principle.  

The final description lengths one is interested in comparing in practice are 
\begin{itemize}
    \item $\mathcal{L}_{C}^{(\text{global})}(\hat G^{(b)})$ (minimum of Eq.~\ref{eq:DLglobalfromBayes})
    \item $\mathcal{L}_{C}^{(\text{local})}(\hat G^{(b)})$ (minimum of Eq.~\ref{eq:DLlocalfromBayes})
    \item $\mathcal{L}_{M}^{(\text{global})}(\hat G^{(b)})$ (minimum of Eq.~\ref{eq:DLglobalmicro})
    \item $\mathcal{L}_{M}^{(\text{local})}(\hat G^{(b)})$ (minimum of Eq.~\ref{eq:DLlocalmicro})
\end{itemize}
where $\hat G^{(b)}$ is the backbone minimizing the corresponding objective for each expression. The objective giving the lowest description length provides the most parsimonious summary of the data $G$ and should be the preferred backboning model.

It is also useful to construct an \emph{inverse compression ratio} to determine the extent to which we can compress a network using its MDL-optimal backbones relative to a na\"ive encoding where we do not use any backbone (equivalent mathematically to setting $G^{(b)}=\{\}$ as the empty graph). For the experiments in Sec.~\ref{sec:results} we will use the microcanonical objectives of Eq.~\ref{eq:DLglobalmicro} and Eq.~\ref{eq:DLlocalmicro}. (These are, as shown in Appendix~\ref{appendix:equivalence}, asymptotically equivalent to the canonical backboning models with geometric weights.) We will compare the levels of compression for the two methods using the inverse compression ratio
\begin{align}\label{eq:compression-ratio}
\eta^{(\text{global/local})} &= \frac{\mathcal{L}_{M}^{(\text{global/local})}(\hat G^{(b)})}{\text{max}[\mathcal{L}_{M}^{(\text{global})}(\{\}),\mathcal{L}_{M}^{(\text{local})}(\{\})]},
\end{align}
where global/local indicates the model of interest. Eq.~\ref{eq:compression-ratio} is a useful ratio in practice because: 
\begin{enumerate}
    \item It is normalized in $[0,1]$ even for very small networks where $\mathcal{L}_{M}^{(\text{global})}(\{\})$ and $\mathcal{L}_{M}^{(\text{local})}(\{\})$ may differ considerably.
    \item It is proportional to the final description length values, so that $\eta^{(\text{global})} > \eta^{(\text{local})}$ only when the global method compresses better than the local method in absolute terms (and vice versa for $\eta^{(\text{global})} < \eta^{(\text{local})}$).
\end{enumerate}

\begin{figure*}
    \centering
    \includegraphics[width=0.9\textwidth]{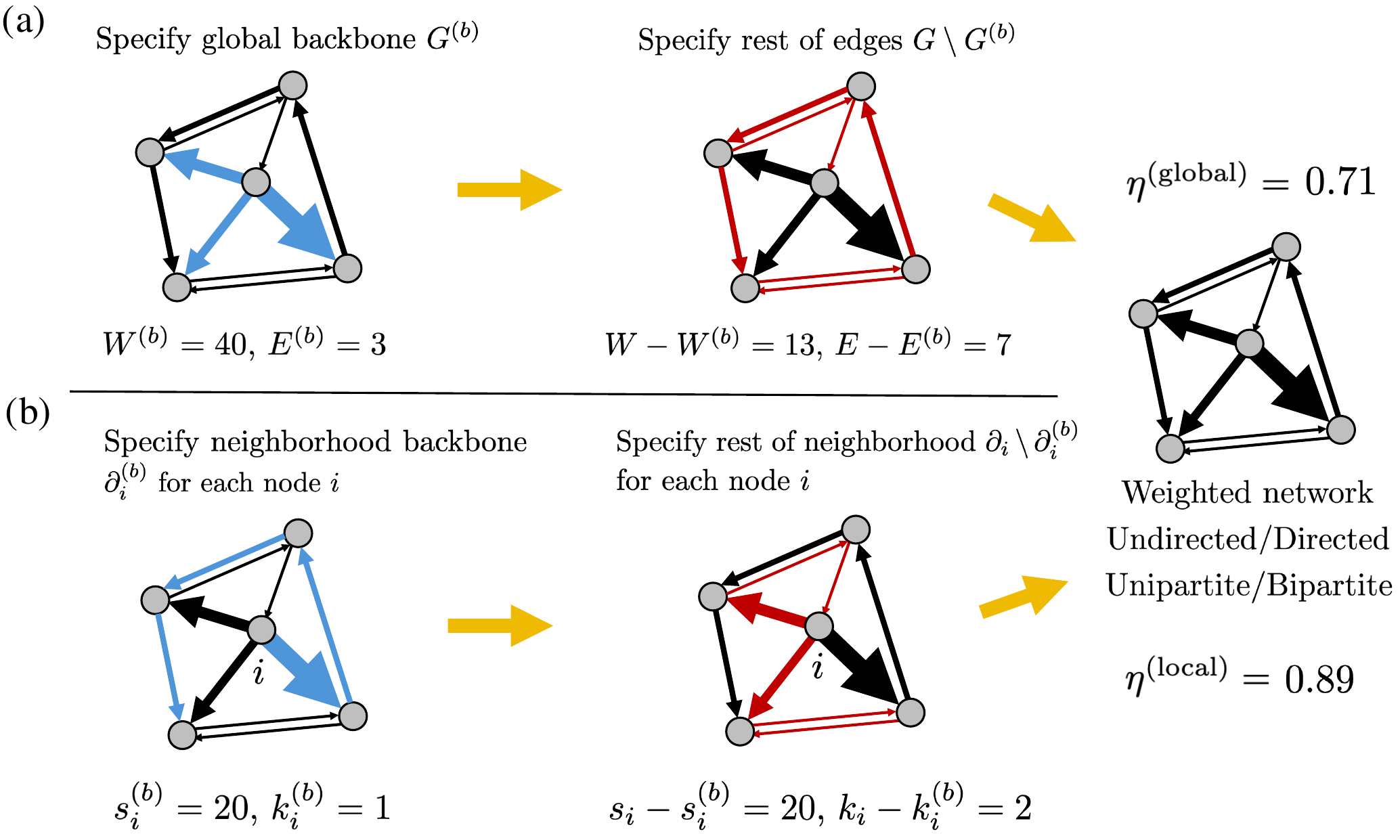}
    \caption{
    \textbf{Global and local MDL backbones for a small example network.}
    \textbf{(a)} Schematic of the global backbone objectives (Eq.~\ref{eq:DLglobalfromBayes} and Eq.~\ref{eq:DLglobalmicro}), with the total weight and number of edges indicated for the backbone $G^{(b)}$ (left) and remaining edges (right) of a network $G$. For the canonical objective (Eq.~\ref{eq:DLglobalfromBayes}), edge weights $w_e$ are sampled according to $w_e\sim P(\cdot \vert \bm{\theta}_{b_e})$, where $b_e\in \{0,1\}$ is an indicator variable for whether or not $e$ is in the backbone. Meanwhile, for the microcanonical objective (Eq.~\ref{eq:DLlocalmicro}), edge weights are distributed uniformly in the backbone and non-backbone given the constraints imposed by $E^{(b)},W^{(b)}$, which are sampled uniformly given the total weight $W$ and number of edges $E$ in the network. \textbf{(b)} Schematic of the local backbone transmission objectives (Eq.~\ref{eq:DLlocalfromBayes} and Eq.~\ref{eq:DLlocalmicro}), applied to out-neighborhoods $\partial_i$ for each node $i$. The total strength $s_i^{(b)}$ and degree $k_i^{(b)}$ are indicated for the backbone neighborhood $\partial^{(b)}_i$ around an example node $i$. For both panels, backbone edges are highlighted in blue, while non-backbone edges are highlighted in red. The local network backbone $G^{(b)}$ is the union $G^{(b)}=\cup_{i=1}^{N}\partial^{(b)}_i$ of the neighborhood backbones. For this example network, the global method provides a more compressive backbone, giving an inverse compression ratio (Eq.~\ref{eq:compression-ratio}) of $\eta^{(\text{global})}=0.71$ versus $\eta^{(\text{local})}=0.89$ for the local backbone.
    }
    \label{fig:diagram}
\end{figure*}

Fig.~\ref{fig:diagram} shows a diagram of both the global and local backboning methods for a small synthetic example network, along with the inverse compression ratios for the microcanonical models. Code implementing these algorithms can be found in the PANINIpy package for nonparametric network inference \cite{kirkley2024paninipy}.

\section{Results}
\label{sec:results}

\subsection{Synthetic Backbone Reconstruction}
\label{sec:reconstruction}

\begin{figure*}
    \centering
    \includegraphics[width=0.9\textwidth]{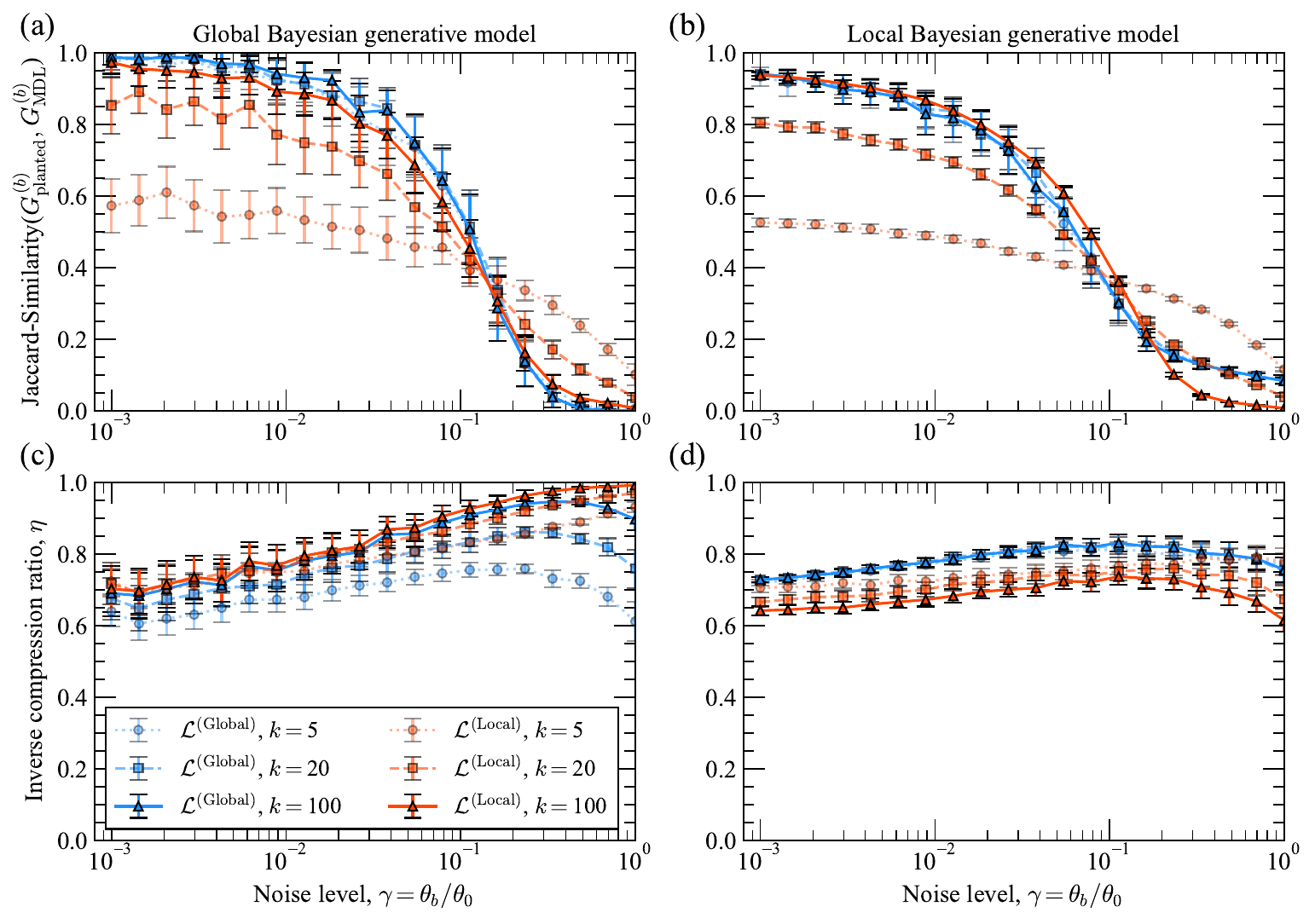}
    \caption{
    \textbf{Reconstruction of planted backbone structure in synthetic network data.}
    \textbf{(a)} Network Jaccard similarity (Eq.~\ref{eq:jaccard}) between the planted and inferred backbones $G^{(b)}_{planted}$ and $G^{(b)}_{MDL}$, using the microcanonical global and local MDL objectives in Eq.s~\ref{eq:DLglobalmicro} and~\ref{eq:DLlocalmicro} respectively. The difficulty of the reconstruction task is varied by changing the noise parameter $\gamma=\theta_1/\theta_0$---the ratio of the means for geometric weight distributions in the non-backbone and backbone---as well as the degree $k$ of the nodes in the underlying random regular directed graph with $N=100$ nodes. The synthetic networks in this panel are drawn from the global canonical model in Eq.~\ref{eq:jointcanonicalglobal} with geometric weights. \textbf{(b)} Network Jaccard similarity between the planted and inferred backbones, for networks generated with the local canonical model of Eq.~\ref{eq:jointcanonicallocal} with geometric weights. \textbf{(c)} Inverse compression ratio (Eq.~\ref{eq:compression-ratio}) for the same set of experiments as in panel (a). \textbf{(d)} Inverse compression ratio for the same experiments as in panel (b). Markers indicate averages over 100 simulations and error bars represent one standard error in the mean.}
    \label{fig:reconstruction}
\end{figure*}

As a first set of experiments to test our backboning methods, we examine the ability of these methods to reconstruct planted backbone structure in synthetic data simulated from the canonical generative models described in Sec.~\ref{sec:canonical}. In the experiments we first generate an unweighted, directed graph with a fixed number of out-edges $k$ for each node $i$. The focus of the MDL backboning methods is on the edge weight distributions, and degree fluctuations simply provide noise in the reconstruction results so degrees are set to be uniform across nodes. We then generate the weights for this network using the global and local canonical generative schemes described in Sec.~\ref{sec:canonical}, with the priors $P(\pi_b)$ and $P(\theta_0)$ (for the global generative model) or $P(\theta_0(i))$ (for the local generative model) uniform over the range $[0,1]$ and $P(w_{e}\vert \theta_{b_{e}})$ or $P(w_{e}\vert \theta_{b_{e}}(i))$ set to a geometric distribution. As a measure of how ``noisy'' the backbone is---in other words, how similar the backbone weights and non-backbone weights are---we fix a parameter $\gamma=\theta_1/\theta_0$ for each simulation which is equivalent to the ratio of the mean non-backbone edge weight and mean backbone edge weight. We set $\theta_1=\gamma\theta_0$ for each simulation to fix the ratio of the mean weight, with the exact mean weights $1/\theta_1$ and $1/\theta_0$ (and consequently the weight variances) varying for each trial based on the sampled value $\theta_{0}\sim \text{Uniform}([0,1])$.

We test the reconstruction and compression capability for the global and local MDL backboning methods in Sec.~\ref{sec:microcanonical} using two measures. The first is the Jaccard similarity index between the planted backbone $G^{(b)}_{\text{planted}}$ generated from the model and the inferred backbone $G^{(b)}_{\text{MDL}}$ for each method. Since the networks $G$ are represented as edge sets, the Jaccard similarity index between a network $G_1$ and a network $G_2$ can be computed as
\begin{align}\label{eq:jaccard}
\text{Jaccard-Similarity}(G_1,G_2) = \frac{\abs{G_1 \cap G_2}}{\abs{G_1 \cup G_2}}.    
\end{align}
The Jaccard similarity in this case tells us how well the global and local MDL backboning algorithms can recover planted structure from their own (approximate canonical) generative models, and falls in the range $[0,1]$ with $0$ indicating an inferred empty or complete graph and $1$ indicating perfect recovery. There are many alternative suitable measures for computing the graph similarity such as the network mutual information \cite{felippe2024network}, but in practice they give the same qualitative trends as the Jaccard similarity for this example. The second measure we use to evaluate the methods is the inverse compression ratio $\eta$ (Eq.~\ref{eq:compression-ratio}), which tells us how well the methods compress the synthetic network data. 

Fig.~\ref{fig:reconstruction} shows the results of our reconstruction experiments, which were run for random directed regular graphs with $N=100$ nodes, degree $k\in \{5,20,100\}$---allowing self-loops---and varying noise levels $\gamma \in [0,1]$. Both the global (left column of panels) and local (right column of panels) generative models were used, to determine the extent to which each backboning method is robust under model mis-specification. We can see from Fig.~\ref{fig:reconstruction}(a) that, as expected, the reconstruction performance becomes worse as we increase the noise level $\gamma$. At $\gamma\approx 10^{-3}$ we observe near perfect backbone recovery for both methods in the high degree regime, but at $\gamma\approx 1$ we see that neither inferred backbone has any significant shared structure with the planted backbone. We can see that the global method is insensitive to degree $k$ while the local method is sensitive to $k$, since lower $k$ will result in greater weight fluctuations within the node neighborhoods but less significant fluctuations at the global level since $E=500$ even for the lowest degree value ($k=5$) studied. We also see that, as expected, the global MDL method has better performance in the low noise regime than the local MDL method, since it is (approximately) the Bayes-optimal algorithm. However, for higher levels of noise $\gamma$ we see that the local backboning method achieves better performance, which improves for smaller degrees $k$. This is because in this regime the global backboning method tends to infer very sparse backbones, but the local method can still find backbones with a moderate level of connectivity due to considering within-neighborhood weight fluctuations rather than global weight fluctuations. 

In Fig.~\ref{fig:reconstruction}(b) we show the reconstruction performance for the same set of experiments but for the local generative model, which allows the model parameters $\pi_b(i),\theta_0(i),\theta_1(i)$ to vary for each node neighborhood $\partial_i$. This causes fluctuations in the weight distributions across nodes, making it more challenging in general to infer the correct backbone structure. Indeed, we can observe in Fig.~\ref{fig:reconstruction}(b) that both methods exhibit a modest drop in reconstruction performance, with the local MDL method now outperforming the global MDL method in the low noise regime for high degrees. Neither method is capable of perfect reconstruction for low noise levels in this more challenging task, but both can still recover the planted backbone with reasonable accuracy.

We plot the inverse compression ratio (Eq.~\ref{eq:compression-ratio}) versus the noise level $\gamma$ in Fig.~\ref{fig:reconstruction}(c) and Fig.~\ref{fig:reconstruction}(d) for the global and local generative models respectively. We see that for the global generative model the global MDL method outperforms the local MDL method, and for the local generative model the local MDL method outperforms the global MDL method. This is consistent with the global and local methods being approximately Bayes-optimal for the global and local generative models respectively, due to the asymptotic equivalence discussed in Appendix~\ref{appendix:equivalence}. In both cases substantial compression is possible using both methods.

In Appendix~\ref{appendix:optimality}, we demonstrate numerically the optimality of the greedy algorithm in similar reconstruction tests on much smaller networks where exact enumeration over backbones is possible, to support the calculations in Sec.~\ref{sec:optimization}.

The results in Fig.~\ref{fig:reconstruction} demonstrate the capability for these methods to identify meaningful backbone structure and provide considerable compression of network data in the presence of noise and model mis-specification. In the next section we compare the MDL methods against existing methods on synthetic data with various levels of planted global and local homogeneity in the edge weights.

\subsection{Comparison on Synthetic Networks}
\label{sec:synthetic}

\begin{figure*}
    \centering
    \includegraphics[width=\textwidth]{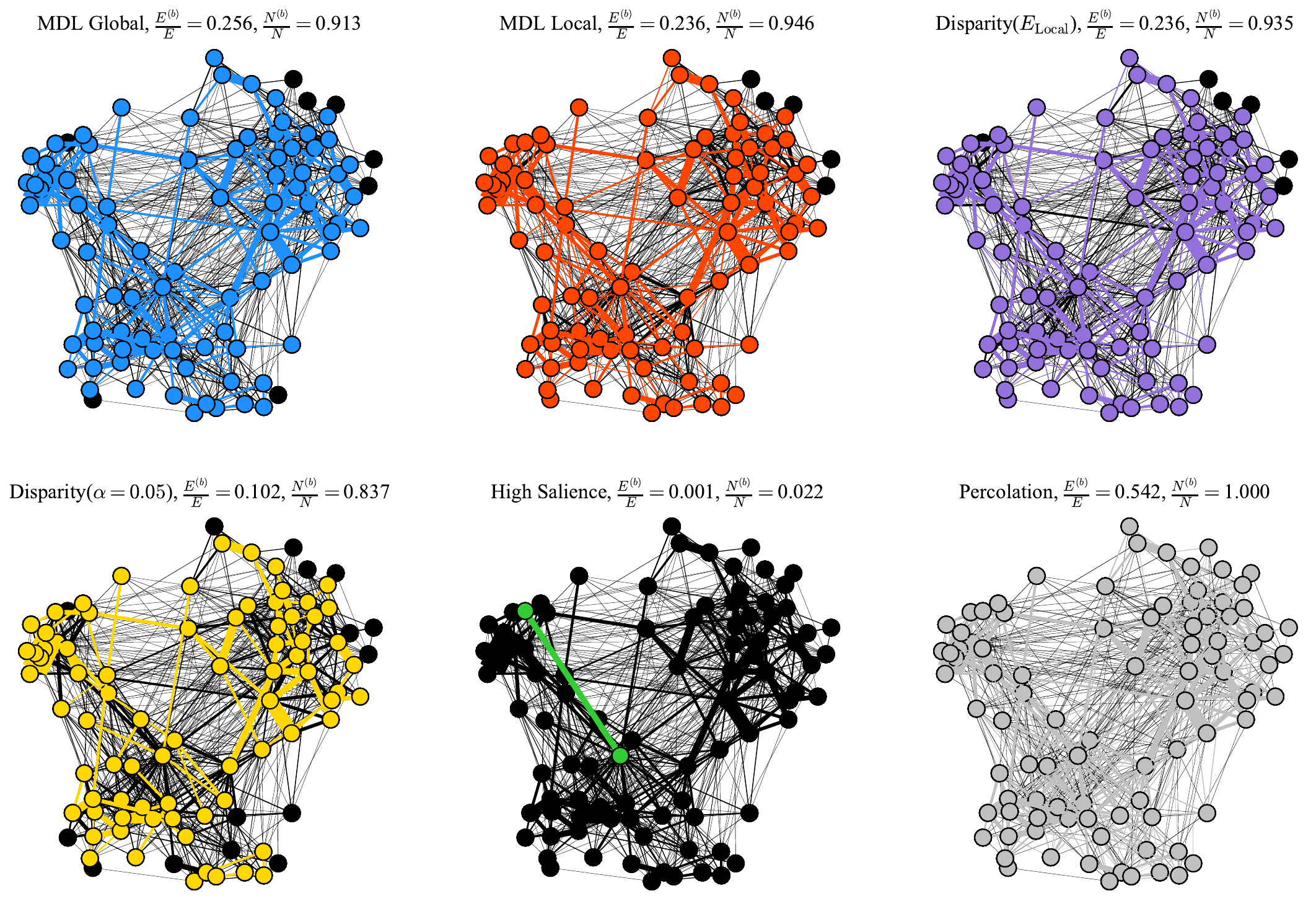}
    \caption{
    \textbf{Six backboning methods applied to a network of workplace contacts.} 
    All six methods used for the tests in Sec.~\ref{sec:results} were applied to the workplace contacts network of \cite{genois2015data}, obtained from the Netzschleuder repository \cite{netz} along with the other networks described in Sec.~\ref{sec:real}. Edges in the backbone are highlighted in color, while all other edges are black. Edges are scaled proportionally to weight. The fraction of edges $E^{(b)}/E$ and nodes $N^{(b)}/N$ retained in the backbone are listed alongside each backboning method above the corresponding plot. The backbone densities and node connectivity (in terms of number of edges retained and number of non-isolated nodes in the backbone) observed for this network are fairly consistent with those observed for other real networks. The percolation backbone has the highest connectivity at the cost of a higher density; the disparity filter with $\alpha=0.05$ and the high salience skeleton have the lowest density at the cost of the lowest connectivity; and the three methods of the top row (global MDL, local MDL, and disparity filter with $E$ equal to the local MDL method) retain most of the node connectivity with lower edge densities.
    }
    \label{fig:real-plots}
\end{figure*}

\begin{figure*}
    \centering
    \includegraphics[width=\textwidth]{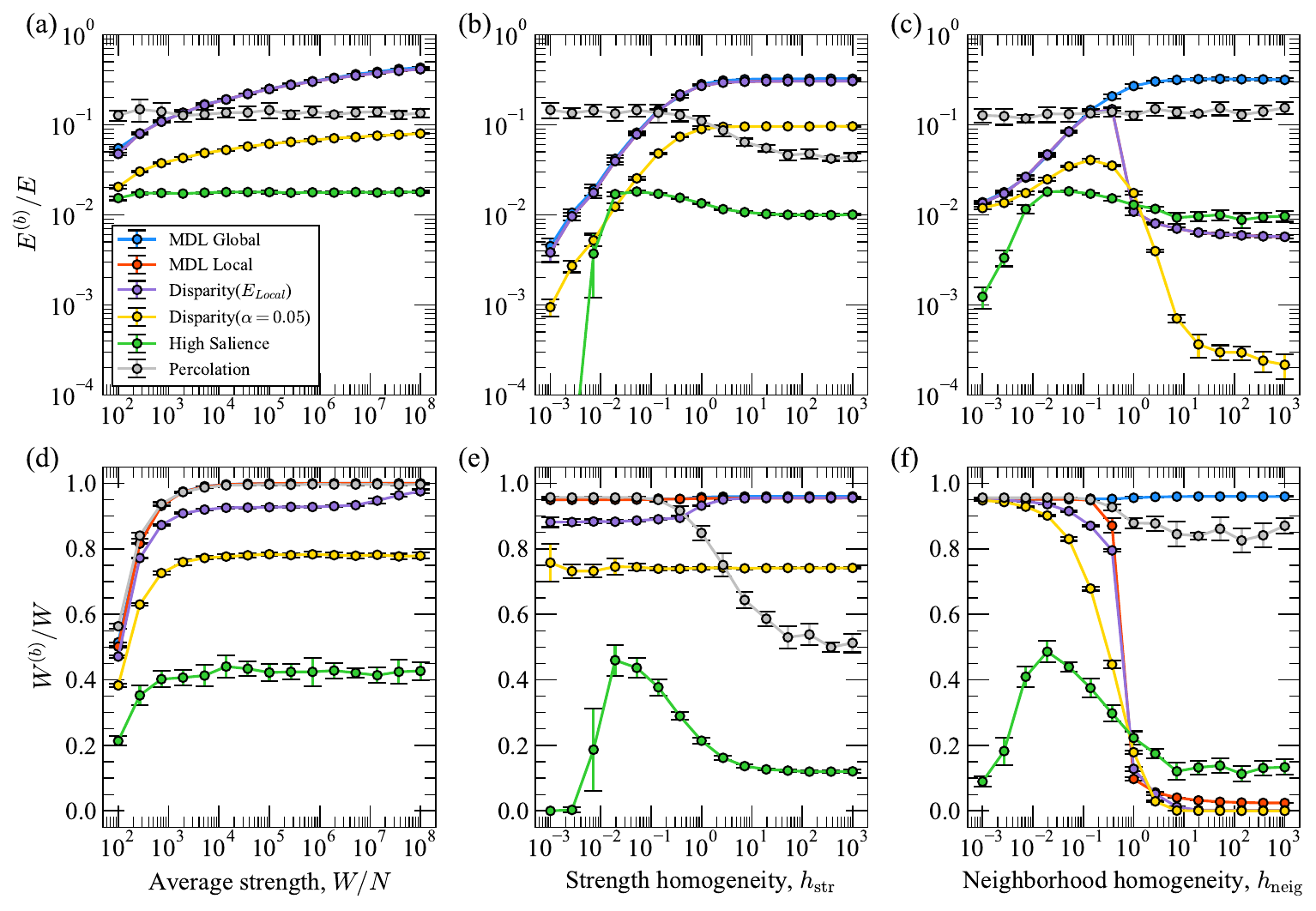}
    \caption{
    \textbf{Concentration of weight along different backbones.} Top row: Fraction of edges retained in the backbone, $E^{(b)}/E$, versus the \textbf{(a)} average node strength $W/N$, \textbf{(b)} level of homogeneity $h_{\text{str}}$ in the node strength distribution, and \textbf{(c)} level of homogeneity $h_{\text{neig}}$ in the weights within each node neighborhood. Bottom row: Panels \textbf{(d)}-\textbf{(f)} plot the fraction of weight retained in the backbone, $W^{(b)}/W$, against the same parameters. Markers indicate averages over 10 simulations from the synthetic network model described in Sec.~\ref{sec:synthetic}, and error bars represent one standard error in the mean.}
    \label{fig:synthetic-EW}
\end{figure*}

\begin{figure*}
    \centering
    \includegraphics[width=\textwidth]{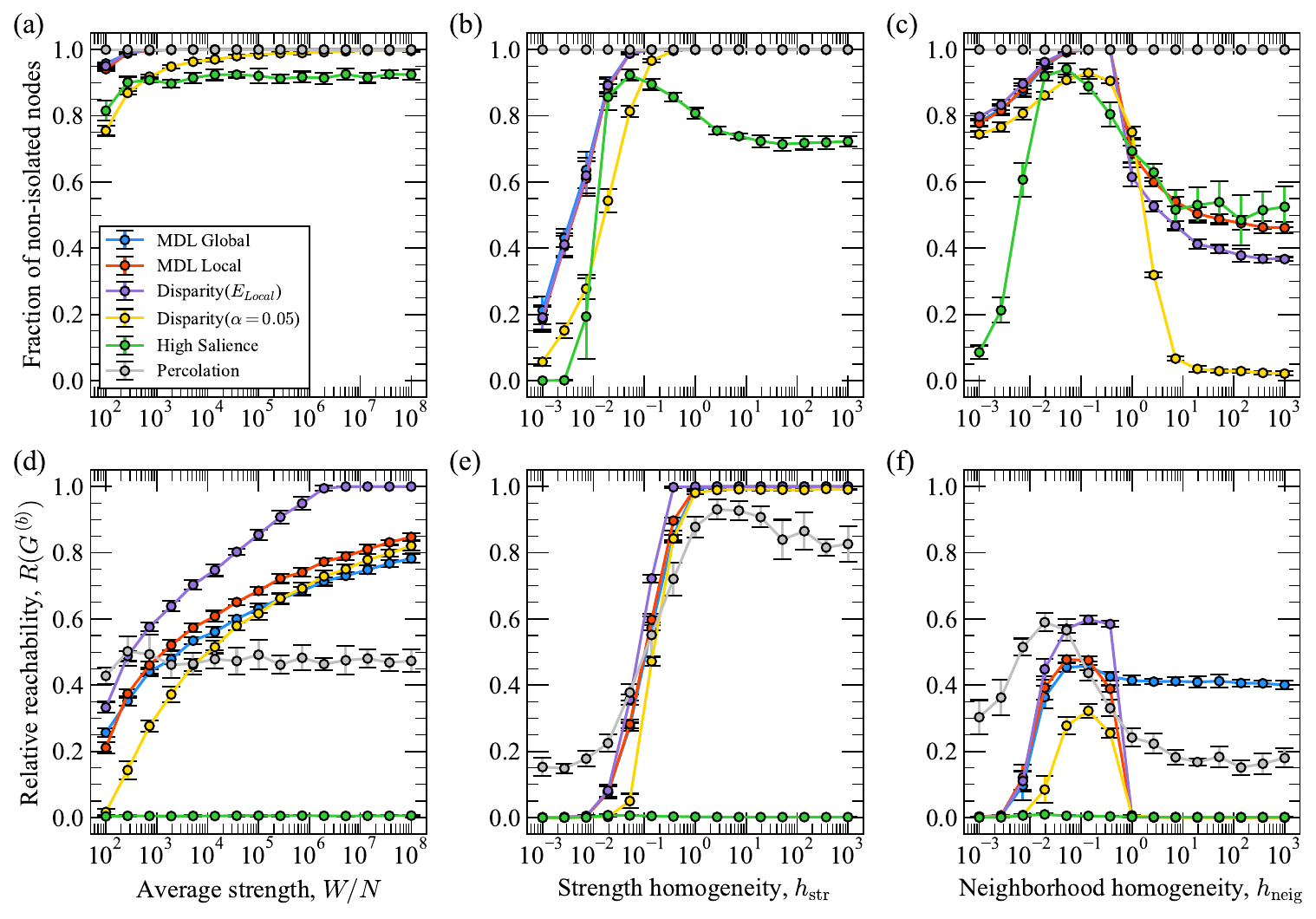}
    \caption{
    \textbf{Global connectivity of different backbones.}
    Top row: Fraction of non-isolated nodes retained in the backbone versus the \textbf{(a)} average node strength $W/N$, \textbf{(b)} level of homogeneity $h_{\text{str}}$ in the node strength distribution, and \textbf{(c)} level of homogeneity $h_{\text{neig}}$ in the weights within each node neighborhood. Bottom row: Panels \textbf{(d)}-\textbf{(f)} plot the relative reachability (Eq.~\ref{eq:reachability}) against the same parameters. Markers indicate averages over 10 simulations from the synthetic network model described in Sec.~\ref{sec:synthetic}, and error bars represent one standard error in the mean.}
    \label{fig:synthetic-RN}
\end{figure*}

\begin{figure*}
    \centering
    \includegraphics[width=\textwidth]{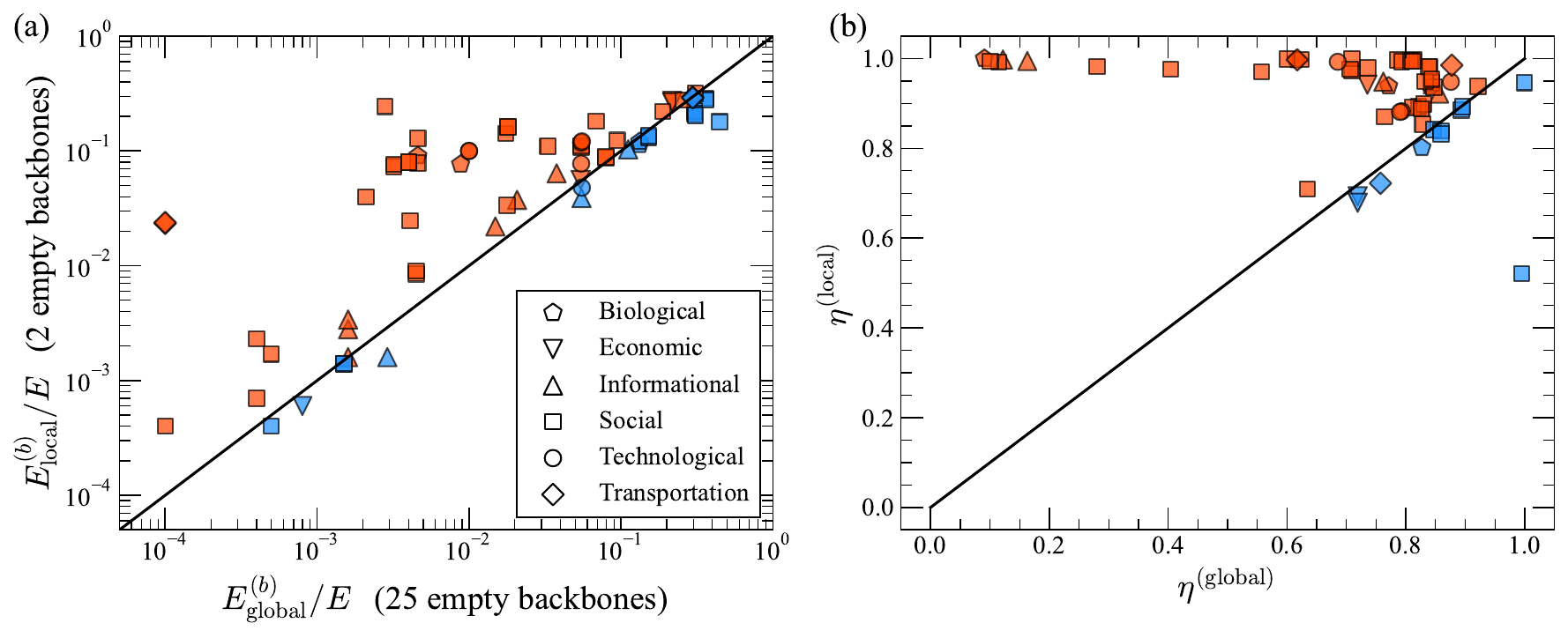}
    \caption{
    \textbf{Global and local backbones across a real-world network corpus.} 
    Global and local MDL backbones were inferred using the objectives of Eq.~\ref{eq:DLglobalmicro} and Eq.~\ref{eq:DLlocalmicro} respectively, for the set of real networks described in Sec.~\ref{sec:real}. We plot \textbf{(a)} the fraction of initial edges retained in the local MDL backbone and global MDL backbone and \textbf{(b)} the inverse compression ratio (Eq.~\ref{eq:compression-ratio}) for each backbone examined. We only include examples for which both the global and local MDL methods inferred a non-empty backbone, noting the empty backbones in the axis labels. In both panels we plot the line $y=x$ for reference and color the examples for which the global (local) backbone returned a higher value with blue (red). We also indicate with different markers the domain of each example network. 
    }
    \label{fig:real-global-v-local}
\end{figure*}

The MDL methods we propose in this paper are fully nonparametric, in contrast to many existing methods which require the specification of the desired number of edges $E^{(b)}$ for the backbone or require a significance level at which to retain edges under some null model for the weights \cite{serrano2009extracting,foti2011nonparametric,marcaccioli2019polya,yassin2023evaluation}. This makes direct comparison with existing methods challenging, since it is unclear how to choose free parameters for methods that require them. 

We therefore select a few popular existing methods for global and local backboning for comparison, with parameters fixed to common values. The first method we use for comparison is the Disparity Filter \cite{serrano2009extracting}, which is a principled inference-based method for local backboning that selects edges below a pre-specified significance level $\alpha$ within each node neighborhood for the backbone. The null model used in the Disparity Filter is one in which the weight assignment is uniform conditioned on the node degree and strength. For more direct comparison with our own local method, for one baseline we fix $E^{(b)}=\abs{G^{(b)}_{\text{local}}}$ as the number of edges to retain in the Disparity Filter and set $\alpha$ to obtain the desired number of edges. For the other Disparity Filter baseline, we set the p-value threshold to the common value of $\alpha=0.05$. We constrain the method to focus on out-neighborhoods for these synthetic examples. 

For the first global backboning baseline, we use the High Salience Skeleton \cite{grady2012robust}, which computes the saliency of a link based on its occurrence frequency in the shortest path trees rooted at each node, giving a succinct global view of edge importance. It is observed that the saliency distribution of links in real networks is highly bimodal, so the skeleton is not sensitive to the specific choice of the saliency cutoff for the backbone, which is chosen to be the center of the saliency distribution ($0.5$). The same methodology is applied here for the baseline. Due to its computation of the shortest path tree for each node, the High Salience Skeleton can become too computationally expensive for large networks, so for $N>10000$ we randomly sample the shortest path trees for $10000$ nodes to estimate the link saliency. For the other global backboning baseline, we use the weighted percolation threshold method \cite{li2007transport,haimovici2013brain}, which consists of adding the edges to the network in decreasing order of weight until the network forms a fully connected component. In Fig.~\ref{fig:real-plots} we plot the results of all six measures applied to the workpace contacts dataset of \cite{genois2015data}, obtained using the procedures described in Sec.~\ref{sec:real}.

The synthetic networks used for the comparison experiments were generated using the following procedure. First, as in Sec.~\ref{sec:reconstruction}, an unweighted network of $N=1000$ nodes is generated as a $k$-regular random directed graph with the specified out-degree $k=50$. Given an input parameter $W$ for the total edge weight, all edges are then assigned weight $1$ and the excess weight $W-Nk$ is distributed across the nodes' out-neighborhoods using a symmetric Dirichlet-Multinomial distribution with concentration parameter $h_{\text{str}}$. This allows us to tune the level of homogeneity in the out-strengths $\{s_i\}$ across nodes with a single parameter $h_{\text{str}}$: $h_{\text{str}}\to 0$ places all the excess edge weight in a single out-neighborhood $\partial_i$ to give a highly heterogeneous strength distribution, and $h_{\text{str}}\to \infty$ distributes all the weight equally across nodes to give a perfectly homogeneous strength distribution. We then distribute the total excess edge weight $s_i-k$ among the out-edges in the neighborhood $\partial_i$ using another symmetric Dirichlet-Multinomial distribution, this time with concentration parameter $h_{\text{neig}}$. $h_{\text{neig}}\to 0$ places all the excess edge weight on a single edge $w_{ij}$ to give a highly heterogeneous weight distribution in each neighborhood, and $h_{\text{neig}}\to \infty$ distributes all the weight equally across edges to give a perfectly homogeneous weight distribution within the neighborhoods. The synthetic networks therefore depend on the four parameters $\{N,W,h_{\text{str}},h_{\text{neig}}\}$, which we vary in our experiments.

We compare backboning methods using multiple indicators, similar to the methodology used in \cite{yassin2023evaluation}. We compare the fraction of edges $E^{(b)}/E$ and total weight $W^{(b)}/W$ appearing in the backbone to get a sense of how sparse the inferred backbones are with each method.
We also compute the inverse compression ratio (Eq.~\ref{eq:compression-ratio}), using the global description length (Eq.~\ref{eq:DLglobalmicro}) for the global methods (global MDL, High Salience Skeleton, Percolation) and the local description length (Eq.~\ref{eq:DLlocalmicro}) for the local methods (local MDL, Disparity Filters) to see how well each method compresses the data. To examine the extent to which the backbones maintain global connectivity we compute the fraction of nodes $N^{(b)}/N$ with non-zero degree in the backbone and the relative \emph{reachability} $R$, defined by
\begin{align}\label{eq:reachability}
R(G^{(b)}) = \frac{\text{\# pairs $i,j$ with directed path $i\to j$ in $G^{(b)}$}}{\text{\# pairs $i,j$ with directed path $i\to j$ in $G$}}.    
\end{align}
Since $R(G^{(b)})$ can become computationally prohibitive for large networks, for $N>10000$ we randomly sample a subgraph of $10000$ nodes to estimate the reachability. We also determine the extent to which the strength distribution has been preserved across nodes by computing the Hellinger distance between the strengths $\{s_i\}_{i=1}^{N}$ in the original network $G$ and the strengths $\{s^{(b)}_i\}_{i=1}^{N}$ in each backbone $G^{(b)}$ using 
\begin{align}\label{eq:hellinger}
D_{\text{str}}(G,G^{(b)}) = \sqrt{\frac{1}{2}\sum_{i=1}^{N}(\sqrt{p_i}-\sqrt{q_i})^2},   
\end{align}
where $p_i=s_i/W$ and $q_i=s_i^{(b)}/W^{(b)}$.

In Fig.~\ref{fig:synthetic-EW} we plot the fraction of edges (top row) and weight (bottom row) retained in the different backbones of the different backboning methods versus the synthetic model parameters. Unless otherwise specified as an independent variable, in all simulations we set $N=1000$, $W/N=1000$, $k=50$, and $h_{\text{str}}=h_{\text{neig}}=0.1$. There are a few major trends we can observe here. The first is that when $h_{\text{neig}}=0.1$ is held constant at a moderate value (first two columns in Fig.~\ref{fig:synthetic-EW}), both MDL methods consistently retain a similar number of edges (an average of around $20\%$ of all edges), which is the highest among all methods for $W/N\gtrsim 10^3$. We can also see that the High Salience Skeleton consistently retains the fewest edges (an average of around $2\%$ of all edges) in these cases, while the Disparity Filter with an $\alpha=0.05$ significance level and the percolation threshold find an intermediate number of backbone edges (an average of around $8\%$ and $12\%$ of all edges respectively). We also observe that the local MDL method consistently retains a higher total backbone weight $W^{(b)}$ for the same number of edges $E^{(b)}$ as the Disparity Filter when the latter is set to have the same edge count. The three local backboning methods are highly sensitive to variations in the heterogeneity of weights within the node neighborhoods $h_{\text{neig}}$ (panels (c) and (f)), while the global MDL method and the High Salience Skeleton are less sensitive to these fluctuations and the percolation method is highly insensitive to the fluctuations. The local backboning methods all have a sudden drop in the number of edges and total weight they retain at $h_{\text{neig}} \approx 1$, which is the point at which the within-neighborhood variance in the edge weights drops significantly according to this Dirichlet concentration parameter. This high level of weight homogeneity within neighborhoods leaves little statistical evidence to retain any edges in the backbone, since no subset of weights are significantly higher than the rest. 

While the size of the backbones is an important factor for efficient downstream graph computations, it is also important that the sparsified networks retain their global connectivity. In the top row of Fig.~\ref{fig:synthetic-RN} we plot the fraction of non-isolated nodes in the backbone (i.e. nodes with degree $0$) versus the three model parameters. We can observe that all backbones except the High Salience Skeleton have nearly zero isolated nodes for a large portion of the parameter space in panels (a) and (b), with higher levels of node isolation for very low levels of strength homogeneity $h_{\text{str}}\lesssim 10^{-2}$ due to sparser backbones across the board. In panel (c) we again see strong fluctuations for the local backboning methods---particularly for the Disparity Filter at $\alpha=0.05$---with the MDL method again performing the best out of the local methods and the global MDL method outperforming the High Salience skeleton in terms of node connectivity. The percolation-based method, by definition, has zero isolated nodes for all parameter settings, at the cost of consistently high fractions of edges being retained in this homogeneous weight regime (Fig.~\ref{fig:synthetic-EW}c). 

The bottom row of Fig.~\ref{fig:synthetic-RN} shows similar connectivity patterns across methods but this time with respect to reachability (Eq.~\ref{eq:reachability}). Here the Disparity Filter has a slight edge over the local MDL method in most experiments for the same number of edges and consistently worse reachability for $\alpha=0.05$. The High Salience Skeleton has exactly zero reachabilty in all experiments, suggesting that although this method highlights important edges for global routing it may tend to construct backbones that are too sparse to maintain global connectivity in the network. The percolation backbone, despite enforcing a single weakly connected component, does not fully retain directed connectivity as required by $R(G^{(b)})$, scoring lower than other methods in most tests. All methods except the global MDL and percolation methods have poor connectivity in the regime of high neighborhood weight homogeneity, becoming too sparse to have a high reachability value when there is little signal to distinguish the neighborhood backbones.

In Fig.~\ref{fig:synthetic-CRD} in Appendix~\ref{appendix:synthetic}, we plot a few additional metrics against the synthetic model parameters, to understand how compressive each method is and how well each preserves the strength distribution of the original network. We also plot the sizes of the inferred backbones versus the number of nodes in the original network in the same Appendix (Fig.~\ref{fig:synthetic-app-N}), finding a very weak relationship with this model parameter. Finally, we show the runtime of some different backboning methods versus the network size, in Fig.~\ref{fig:synthetic-app-runtime}. The slightly superlinear scaling of the MDL runtimes is consistent with the theoretical scaling of $O(E\log E)$ discussed in Sec.~\ref{sec:optimization}. We also observe a substantially higher computational complexity for the High Salience Skeleton due to the computation of the shortest path trees.

\subsection{Comparison on Real Networks}
\label{sec:real}

\begin{figure*}
    \centering
    \includegraphics[width=\textwidth]{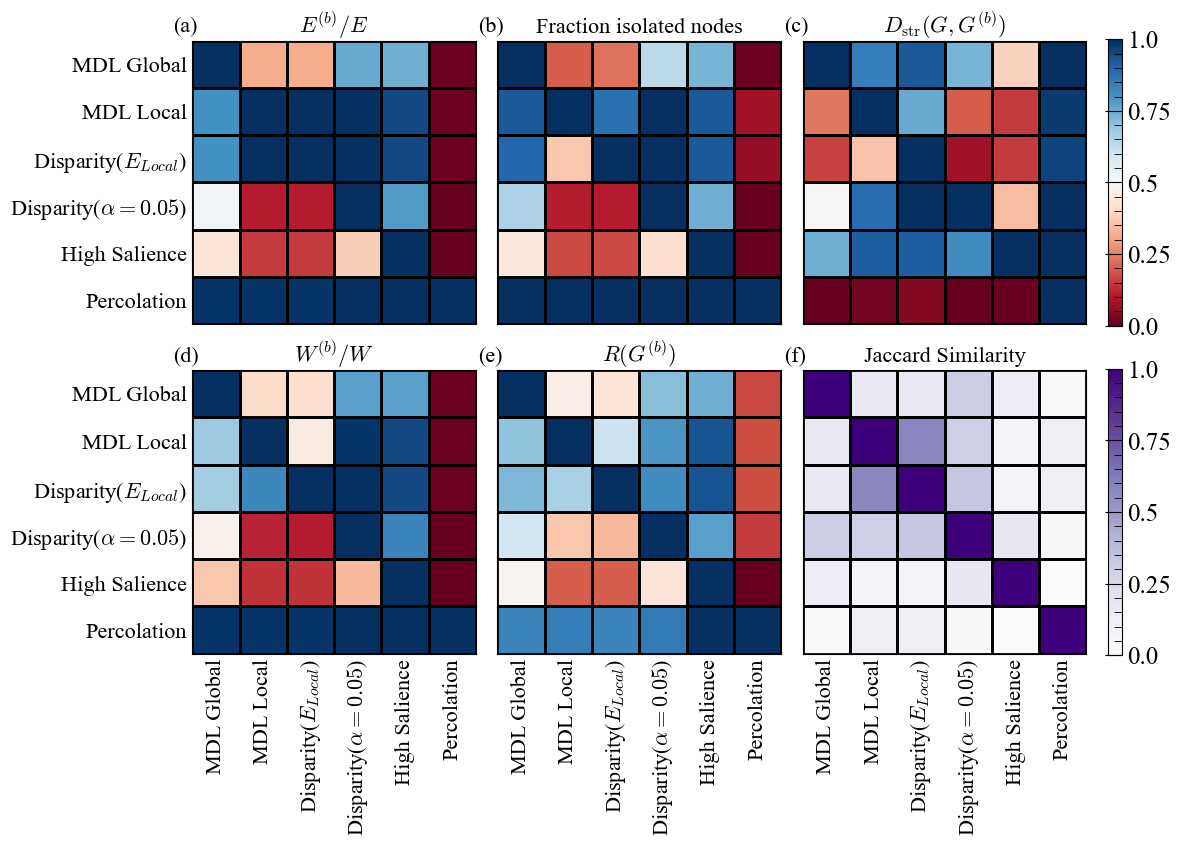}
    \caption{
    \textbf{Comparison of backbone metrics on real networks.} 
     The six backboning methods were compared using the same metrics as in Sec.~\ref{sec:synthetic} on the real networks in the corpus described in Sec.~\ref{sec:real}. In each of the first five panels we plot a matrix $\bm{B}(m)$ for a metric $m$ of interest such that $B_{ij}(m)$ is the fraction of real networks in the corpus for which $m_i\geq m_j$. High values $B_{ij}(m)$ indicate that the metric $m$ was higher for backboning method $i$ than backboning method $j$ in a large portion of real network instances. We plot the matrix $\bm{B}(m)$ for \textbf{(a)} the fraction of edges retained in the backbone, $m=E^{(b)}/E$; \textbf{(b)} the fraction of nodes with degree zero in the backbone; \textbf{(c)} the Hellinger distance (Eq.~\ref{eq:hellinger}) between the strength distributions of the backbone and original network; \textbf{(d)} the fraction of weight retained in the backbone, $m=W^{(b)}/W$; and \textbf{(e)} the relative reachability (Eq.~\ref{eq:reachability}) between the backbone and original graph. In panel \textbf{(f)} we plot the average network Jaccard similarity (Eq.~\ref{eq:jaccard}) between the backbones generated by each pair of methods, with a different color scale to indicate the different interpretation of the heatmap.
    }
    \label{fig:real-comparison}
\end{figure*}

We now compare the six backboning algorithms from Sec.~\ref{sec:synthetic} on a large corpus of real network datasets coming from different application domains. We initially collected 69 networks from the Netzschleuder repository~\cite{netz} by retrieving all weighted networks with edge weights $w_{ij}\geq 1$ and less than $10^7$ edges. Non-integer-valued weights $w_{ij}$ were rounded to the nearest integer for analyses since the microcanonical models were used. As the methods in Sec.~\ref{sec:synthetic} are all adaptable to both directed and undirected networks, we analyzed the 32 directed networks in the corpus using both edge directions separately (in-edges and out-edges) and analyzed the remaining 37 undirected networks with both edge directions simultaneously as described in Sec.~\ref{sec:canonical}. The networks in the final corpus represent a diverse array of domains, and (using the classification provided by \cite{netz}) in total we inferred 54 social network backbones, 11 informational network backbones, 9 biological network backbones, 7 technological network backbones, 6 transportation network backbones, and 5 economic network backbones using each of the six methods. The networks ranged in size from $E=91$ to $E=5,743,258$ with total weight ranging from 
$W=282$ to $W=2,541,576,441$ and average degree ranging from $\expec{k}=1.1$ to $\expec{k}=238.3$.

Our first experiment with this real-world network corpus examines the extent to which the backbones differ between the global and local MDL backboning procedures described in Sec.s~\ref{sec:microcanonical}. In Fig.~\ref{fig:real-global-v-local}(a) we plot the fraction of the original edges retained in the local MDL backbone versus the fraction retained in the global MDL backbone for each of the backbones examined. We only include examples for which both methods found a non-empty backbone, noting the number of violating examples in the axis labels. We can see that the global MDL method tends to find empty backbones in a large fraction of the examples studied (25/92), indicating that this method is frequently not able to compress real network structure any better than a na\"ive edge transmission. On the other hand, only two networks failed to be compressed by the more flexible local MDL method. When both methods give nontrivial backbones, we can see that they return a similar number of edges, with the local MDL method returning more edges in many cases but neither method exceeding $E^{(b)}/E\approx 0.35$. In Fig.~\ref{fig:real-global-v-local}(b) we plot the inverse compression ratios (Eq.~\ref{eq:compression-ratio}) for each method, which indicates that when the global MDL method does compress it does so quite well, even compressing roughly $\approx 90\%$ of the information in the network relative to a naive edge transmission in some cases. We see that the global method compresses better than the local method in many cases, demonstrating that a global threshold can in fact provide an effective backbone for many real networks. Neither panel indicates any clear differentiation among the backbone structure of networks based on domain.

In the next experiment we compare all six backboning methods of Sec.~\ref{sec:synthetic} with respect to the same metrics---specifically, the fraction of edges retained in the backbone; the fraction of nodes with degree zero in the backbone; the Hellinger distance (Eq.~\ref{eq:hellinger}) between the strength distributions of the backbone and original network; the fraction of weight retained in the backbone, $m=W^{(b)}/W$; and the relative reachability (Eq.~\ref{eq:reachability}) between the backbone and original graph. We omit analysis of the inverse compression ratio since trivially the global and local MDL methods will always obtain better compression than their global and local counterparts, and the direct comparison of the two MDL methods is done in Fig.~\ref{fig:real-global-v-local}. Fig.~\ref{fig:real-comparison}(a)-(e) shows the results of these experiments through pairwise comparisons among all pairs of backboning methods for each of the six measures. We can see that in the real networks the methods are roughly ordered like $\text{Percolation}>\text{MDL Local}\approx\text{Disparity}(E_{Local})>\text{MDL Global}>\text{Disparity}(\alpha=0.05)>\text{High Salience}$ when considering the number of edges, the total weight, the fraction of isolated nodes, and the relative reachability of the inferred backbones. Meanwhile, the reverse pattern is observed for the Hellinger distance. MDL Local has a slight edge over $\text{Disparity}(E_{Local})$ with respect to the fraction of isolated nodes and the reachability, while $\text{Disparity}(E_{Local})$ has a slight edge for the Hellinger distance. The percolation backbone performs the best overall in these tests but at the cost of retaining a substantial fraction of edges (an average of $84\%$!). 

In Fig.~\ref{fig:real-comparison}(f) we plot the Jaccard similarity (Eq.~\ref{eq:jaccard}) between the backbones generated by each pair of methods, averaged over all networks in the corpus. We observe very low levels of similarity among all pairs of backbones except for those generated by the $\text{MDL Local}$ and $\text{Disparity}(E_{\text{Local}})$ methods, which have a moderately high average overlap of roughly $0.5$. 

Overall, the results in Sec.~\ref{sec:synthetic} and Sec.~\ref{sec:real} indicate that the local MDL method and the Disparity Filter set to the same number of backbone edges as the local MDL method tend to best preserve the original network structure in many cases. Meanwhile, the global MDL method often finds empty backbones and does not compress relative to a naive transmission scheme, but in the instances that it does compress it performs well across the metrics while retaining fewer edges and providing even better compression than the local method. Setting the Disparity filter to the typical significance level of $\alpha=0.05$ tends to be too conservative in some real networks, suggesting that this parameter needs to be carefully tuned in practice for different applications. The High Salience Skeleton is quite conservative in its backbone estimation, as it highlights only a few links that are the most significant for global network routing. And finally, the percolation backbone tends to preserve connectivity and weight heterogeneity but does not provide substantial sparsification, retaining a majority of the edges in most networks. 


\subsection{Comparison of Spreading Dynamics on Network Backbones}
\label{sec:MPsims}

\begin{figure*}
    \centering
    \includegraphics[width=\textwidth]{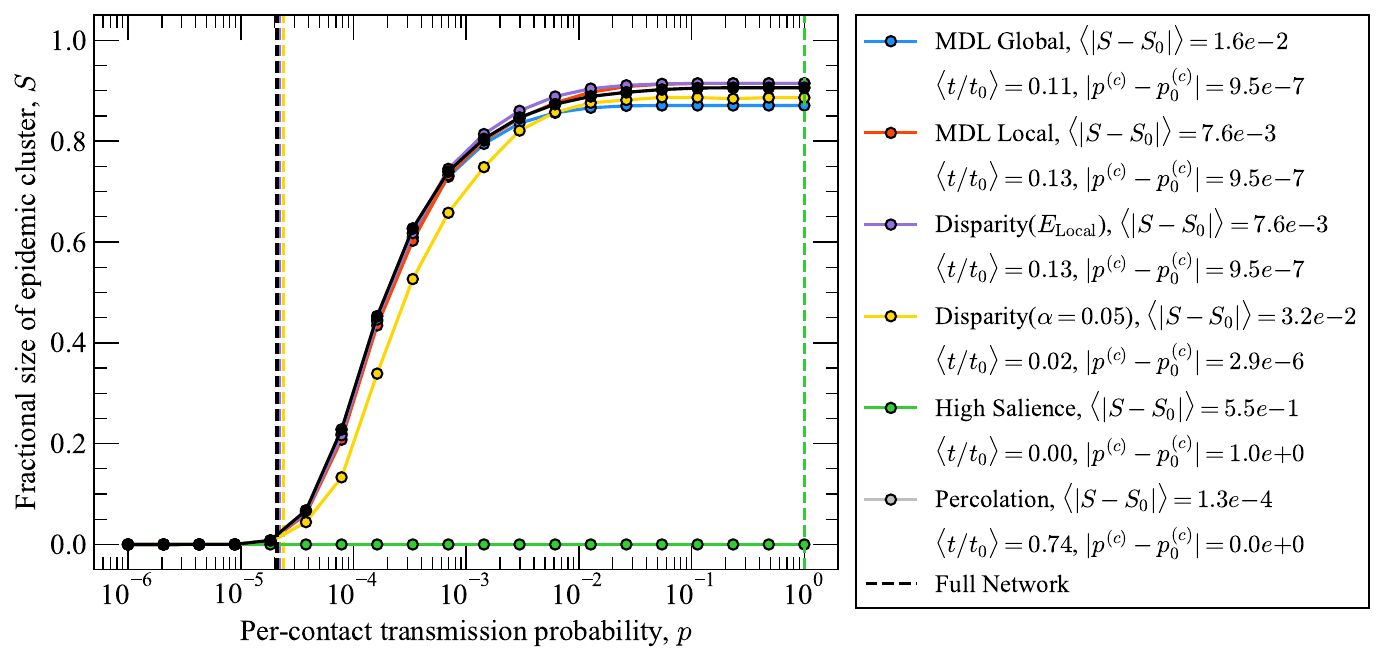}
    \caption{
    \textbf{Speeding up epidemic calculations with network backbones.} 
     Fractional size of the epidemic cluster (Eq.~\ref{eq:S}) versus the per-contact transmission probability $p$ for the message passing system of Eq.s~\ref{eq:uij} and~\ref{eq:si}. The percolation threshold estimated using Eq.~\ref{eq:stability} for each network backbone and the full network (black) are shown with dotted vertical lines. The average error in $S$, runtime savings, and error in the percolation threshold (as described in Sec.~\ref{sec:MPsims}) are shown in the right panel. 
    }
    \label{fig:spreading}
\end{figure*}

As an example downstream application of our backboning methods, we examine their performance against baselines for sparsifying networks while retaining dynamical properties relevant for spreading and percolation calculations. Specifically, we adapt a standard method for estimating the epidemic threshold from long-term transmission probabilities among each pair of nodes \cite{newman2018networks} to the weighted network case, and study how accurate and computationally expensive each backbone is for this calculation on an example network. 

In the original model discussed in \cite{newman2018networks}, there is a single parameter $\phi$ characterizing the probability that a node $i$ has strong enough contact with any other node $j$ for the disease to get passed from $j\to i$ if $j$ ends up with the disease. This is analogous to the occupation probability in a bond percolation process. In our weighted network adaptation, we consider the parameter $\phi(w_{ij})$ characterizing this probability to be a monotonic function of the weight $w_{ij}$ of the edge between nodes $i$ and $j$. (For simplicity, we restrict our analyses to undirected networks.) This mechanism corresponds to $i$ and $j$ having a greater connectivity and potential to transmit the disease to each other when they have a higher weight edge between them. Specifically, we consider a process in which the weight $w_{ij}$ can be interpreted as the frequency of contact between $i$ and $j$, and that there is a probability $p$ for the disease to spread independently from each contact, so that $\phi(w_{ij})=1-(1-p)^{w_{ij}}$. 

By varying $p\in [0,1]$, we can find a percolation transition in this model in which an extensive giant component forms and an epidemic can spread to the entire network. We can compute this threshold using a method analogous to that of \cite{karrer2014percolation} in which a constant occupation probability is considered. Letting $u_{i\to j}$ be the probability that node $i$ is not connected to giant component through node $j$---in other words, the probability that $j$ does not contribute to the epidemic through passing the disease to node $i$---we have 
\begin{align}\label{eq:uij}
u_{i\to j} = 1-\phi(w_{ij})+\phi(w_{ij})\prod_{k\in \mathcal{N}_{j\setminus i}}u_{j\to k},
\end{align}
where $\mathcal{N}_{j\setminus i}$ is the set of nodes attached to $j$ other than node $i$. The probability that $i$ is part of the giant cluster (e.g. epidemic cluster) is then given by
\begin{align}\label{eq:si}
S_i=1-\prod_{j\in \mathcal{N}_i}u_{i\to j},    
\end{align}
where $\mathcal{N}_i$ is the set of nodes connected to $i$, and the fractional size $S$ of the giant cluster is
\begin{align}\label{eq:S}
S = \frac{1}{N}\sum_{i=1}^{N}s_i = 1-\frac{1}{N}\sum_{i=1}^{N}\prod_{j\in \mathcal{N}_i}u_{i\to j}.  \end{align}
Technically, the above calculation is only exact on trees---otherwise, we cannot assume the independence of $j$'s neighbors to write $\prod_{k\in \mathcal{N}_{j\setminus i}}u_{j\to k}$. However, it often gives a good approximation in practice for tree-like sparse networks. (If one is interested in more complex approximations accounting for dependence among node neighbors \cite{yedidia2003understanding,kirkley2021belief}, sparsifying the networks ahead of time through backboning may be necessary, for tractability.)  

Equations~\ref{eq:uij}~and~\ref{eq:si} define a system of message passing equations that can be used to solve for the size of the percolating epidemic cluster of any network $G$ as a function of the per-contact transmission probability $p$. To do this, the probabilities $u_{i\to j}$ are initialized to random values, and Eq.~\ref{eq:uij} is iterated until convergence, at which point $S_i$ can be computed for each node $i$ using Eq.~\ref{eq:si}. 

Fig.~\ref{fig:spreading} shows the results of simulating this message passing system on the reality mining network of student contacts collected from mobile phone data in \cite{eagle2006reality}, for $p$ ranging from $10^{-6}$ to $1$. The system was simulated on the full network (black curve) as well as the backbones constructed using the six methods of Sec.~\ref{sec:synthetic}. To compare these curves, we compute the average absolute error $\expec{\abs{S-S_0}}$ between the values $S$ associated with each backboning method and the full network's values $S_0$. We can observe that the percolation-based backboning method performs the best with respect to this metric, although we will see that it does not provide a substantial computational speedup since it retains a large fraction of the edges. The local MDL method and the disparity filter with $E=E_{\text{Local}}$ perform simililarly under this measure, still providing a good approximation of the full network's giant component size at each $p$, while the global MDL method performs slightly worse, deviating from the desired curve for high values of $p$. The disparity filter with $\alpha=0.05$ performs worse yet, with noticeably large deviations as the giant component grows in size, and the high salience skeleton---with very little retained in the backbone---performs the worst out of the six methods for this task.       

We can notice that at a critical probability $p^{(c)}\in [0,1]$, the system transitions from a fragmented state with many small connected components ($S\approx 0$) to having a single extensive giant connected component ($S\approx 1$). There is a trivial fixed point in the message passing system (Eq.~\ref{eq:uij}) of $u_{i\to j}=1$---in other words, when no one is in percolating cluster ($S=0$). We can therefore find the percolation transition by expanding in $u_{i\to j}=1-\epsilon_{i\to j}$ for small $\epsilon_{i\to j}$ and identifying at which value of $p$ this fixed point is no longer stable. When the trivial fixed point is stable, we have no giant component, and when it is unstable, we have a giant component. (A similar calculation is performed in \cite{karrer2014percolation} for constant occupation probability $\phi$.)

Plugging in $u_{i\to j}=1-\epsilon_{i\to j}$ to Eq.~\ref{eq:uij} and expanding for small $\epsilon_{i\to j}$ gives
\begin{align}\label{eq:stability}
\epsilon_{i\to j} = \phi(w_{ij})\sum_{k\in \partial_{j\setminus i}}\epsilon_{j\to k}
=\sum_{i\to j,l\to k} B^{(\phi)}_{i\to j,l\to k}      \epsilon_{l\to k},
\end{align}
where $B^{(\phi)}_{i\to j,l\to k}=\phi(w_{ij})\delta_{jl}(1-\delta_{ik})$ are entries---indexed by edges $i\to j$ and $l\to k$---in a modified version of the non-backtracking (or Hashimoto) matrix that account for the transmission probabilities. The system will percolate when the above fixed point is unstable, i.e. when $\bm{B}^{(\phi)}$ has a leading eigenvalue of magnitude greater than $1$. We can then identify the value $p^{(c)}$ numerically without the need for extensive message passing simulations by  evaluating the leading eigenvalue of $\bm{B}^{(\phi)}$ at some starting guess $p$ and running a binary search over $p$ until the leading eigenvalue of $\bm{B}^{(\phi)}$ has magnitude close to $1$. The value at which this terminates will be an estimate of $p^{(c)}$. 

In Fig.~\ref{fig:spreading} we can examine how well each backboning method preserves the percolation threshold $p^{(c)}$ of the full network and how much runtime it saves. A good balance of these two factors is ideal in practice, as we desire accuracy while saving computational cost. (For the particular example network studied here, we do not face any computational bottleneck. But this serves as a proof-of-concept for application to much larger networks.) We measure the runtime savings $\expec{t/t_0}$ as the fractional reduction relative to the runtime $t_0$ for the full network, averaged over all evaluations of the binary search for $p^{(c)}$. We then measure the error in the percolation threshold as $\abs{p^{(c)}-p^{(c)}_0}$, where $p^{(c)}_0$ is the threshold for the full network. 

We can see that, while the percolation-based backbone has the most accurate estimation of $p^{(c)}$, it fails to reduce the runtime of the computation substantially, requiring on average $74\%$ of the runtime of the full network calculation to compute the leading eigenvalues. On the opposite extreme, the leading eigenvalue calculations on the high salience skeleton require almost no runtime, but are completely inaccurate and estimate $p^{(c)}=1$. The disparity filter with $\alpha=0.05$ has the next highest error in $p^{(c)}$, though provides runtimes much lower than other methods (other than the high salience skeleton). The global MDL, local MDL, and disparity filter set to $E=E_{\text{Local}}$ find a good middle ground, allowing for a roughly $10\times$ speedup in the calculation while approximating the percolation transition to within $6$ decimal places.

These results together indicate that the MDL backbones are capable of preserving not only structural, but also dynamical properties of networks, despite not explicitly involving dynamics in their formulation.


\section{Conclusion}
In this paper we develop a completely nonparametric framework for inferring the backbone of a weighted network which utilizes the minimum description length (MDL) principle to promote sparsity. Our method is adapted to infer both global and local network backbones, and is generalizable to any mixture of weight distributions over the backbone and non-backbone edges using a flexible Bayesian generative model. We develop fast exact optimization schemes for our global and local MDL backboning objectives that are log-linear in the number of edges, allowing these principled methods to easily scale to networks with millions of edges. We compare our method with existing methods in a range of tasks on synthetic and real network data, showing that the proposed MDL methods are capable of substantially sparsifying a wide variety of networks while retaining meaningful global and local structural characteristics.

There are a number of potential avenues for future work extending our methods. One critical limitation of the proposed method is that it does not apply to unweighted graphs, since the weights are used for compression to identify a meaningful backbone. Extending the MDL framework for backboning unweighted graphs is therefore an important avenue for future work. Additionally, as mentioned in Sec.~\ref{sec:canonical}, our method is generalizable for inferring global and local network backbones under any mixture of weight distributions in the backbone and non-backbone, with a subset of these permitting exact minimization with Algorithms~\ref{alg:global} and~\ref{alg:local}. Here we study a microcanonical model asymptotically equivalent to a canonical model with geometric weights, but in future work it is important to examine other distributions which may provide better compression of networks with different types of weight heterogeneity. One could also fully utilize the posterior distribution over backbones in this Bayesian framework to sparsify networks in a stochastic manner, similarly to some existing sparsification techniques \cite{spielman2008graph}. Both of these extensions may incur a substantial additional computational burden for optimization and sampling compared to the fast greedy optimization performed here, so present new challenges for practical implementation. One could also extend our framework to sparsify hypergraphs or other higher order networks with weight metadata. Finally, it may be possible to extend our method to infer sparse functional network backbones by exploiting regularities in dynamical information such as node or edge states.


\section*{Acknowledgments}
\vspace{-\baselineskip}
The author acknowledges support from the HKU-100 Start Up Fund.


\clearpage
\appendix
\onecolumngrid

\section{Asymptotic Equivalence Between Microcanonical Model and Canonical Model with Geometrically Distributed Weights}
\label{appendix:equivalence}
In this Appendix we demonstrate the approximate equivalence between the microcanonical model of Sec.~\ref{sec:microcanonical} and the canonical model of Sec.~\ref{sec:canonical} when weights are distributed according to a geometric distribution. Letting $P(w\vert \theta)=\theta(1-\theta)^{w-1}$ be a geometric distribution and $P(\theta)=1$ a uniform prior on $[0,1]$ (which is conjugate to $P(w\vert\theta)$), we have that Eq.~\ref{eq:DLglobalfromBayes} can be written as
\begin{align}
\mathcal{L}&_C^{(\text{global})}(G^{(b)})
= \log (E+1) + \log {E\choose E^{(b)}} - \log \int \prod_{e\in G^{(b)}}\theta_1(1-\theta_1)^{w_e-1} d\theta_1  - \log \int \prod_{e\in G\setminus G^{(b)}}\theta_0(1-\theta_0)^{w_e-1} d\theta_0 \nonumber\\
&= \log (E+1) + \log {E\choose E^{(b)}}-\log \int \theta_1^{E^{(b)}}(1-\theta_1)^{W^{(b)}-E^{(b)}} d\theta_1 -\log \int \theta_0^{E-E^{(b)}}(1-\theta_0)^{(W-W^{(b)})-(E-E^{(b)})} d\theta_0 \nonumber\\
&=\log (E+1) + \log {E\choose E^{(b)}}-\log B(E^{(b)}+1,W^{(b)}-E^{(b)}+1) -\log B(E-E^{(b)}+1,(W-W^{(b)})-(E-E^{(b)})+1) \nonumber\\
&=\log (E+1) +\log(W^{(b)}+1) +\log(W-W^{(b)}+1)  + \log {E\choose E^{(b)}}+\log {W^{(b)}\choose E^{(b)}} +\log{W-W^{(b)}\choose E-E^{(b)}} \\
&= \mathcal{L}_M^{(\text{global})}(G^{(b)})
+ \delta, \nonumber
\end{align}
where
\begin{align}
\delta = \log \frac{(W^{(b)}+1)(W-W^{(b)}+1)(W^{(b)})(W-W^{(b)})}{(W-E+1)(E^{(b)})(E-E^{(b)})}
\leq \log \frac{(W/2+1)^2(W/2)^2}{(W-E+1)(E-1)}
\end{align}
for $E^{(b)}\in [1,E/2]$. For $E^{(b)}=0$, we only evaluate the second integral since there are no edges in the backbone $G^{(b)}$, and the resulting expressions differ by $\delta = \frac{(W+1)W}{(W-E+1)E}$. Regardless, the correction scales like $\delta \sim O(\log W+\log E)$, and so we have
\begin{align}
\mathcal{L}_C^{(\text{global})}(G^{(b)}) = \mathcal{L}_M^{(\text{global})}(G^{(b)}) + O(\log W+
\log E),    
\end{align}
and the canonical and microcanonical expressions are asymptotically equal for $W,E\gg 1$. 

Similarly, the microcanonical neighborhood-level description length of Eq.~\ref{eq:neigdlcanonical} is equivalent to the neighborhood-level description length of Eq.~\ref{eq:neigdlmicro} up to a correction of $O(\log s_i+\log k_i)$, making these objectives asymptotically equivalent for $s_i,k_i\gg 1$. However, the full canonical local description length of Eq.~\ref{eq:DLlocalfromBayes} is only equivalent to the corresponding microcanonical expression in Eq.~\ref{eq:DLlocalmicro} up to $-\log P(\bm{s})=\log {N+W-E-1\choose W-E}$ in addition to a correction of $O(N\expec{\log k +\log s})$ which is of a similar size. Despite this discrepancy in the network-level description lengths for the local backboning methods, since the backboning is done for each neighborhood $\partial_i$ separately the microcanonical model is easily capable of inferring planted backbone structure generated from the canonical model (as seen in Fig.~\ref{fig:reconstruction}).

\section{Greedy Optimality in Numerical Experiments}
\label{appendix:optimality}

\begin{figure*}[h]
    \centering
    \includegraphics[width=0.7\textwidth]{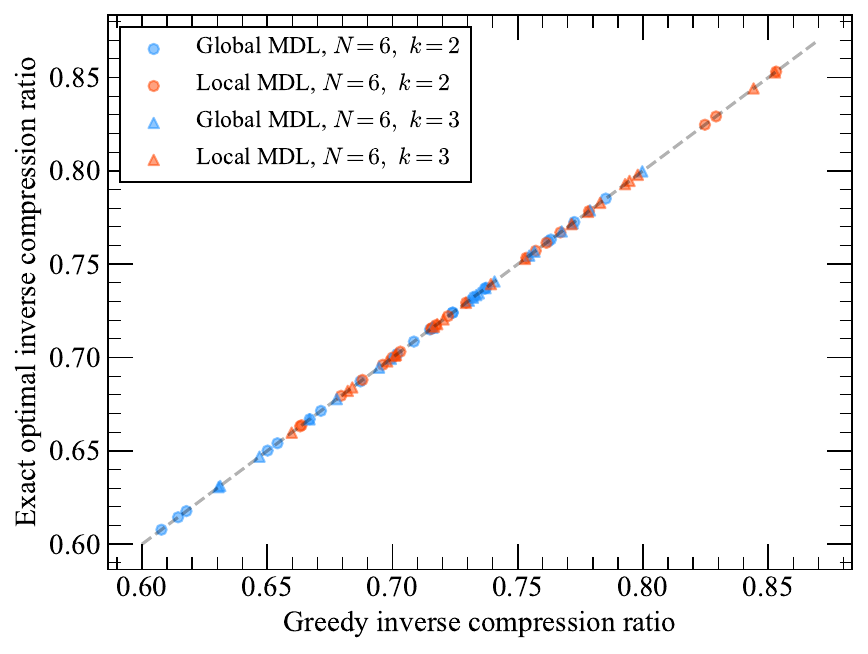}
    \caption{
    \textbf{Optimality of greedy algorithm.}
    Inverse compression ratios for all reconstruction experiments in Fig.~\ref{fig:reconstruction}, performed on networks with $N=6$ and $k\in \{2,3\}$. The results obtained from the greedy methods of Algorithm~\ref{alg:global} and Algorithm~\ref{alg:local} are shown on the x-axis, while the results obtained through exhaustive enumeration over possible backbones are shown on the y-axis. We find that the greedy algorithms give identical results to exact enumeration, as expected, for all cases studied.}
    \label{fig:optimality}
\end{figure*}

Here we plot the results of reconstruction experiments using the same generative model as in Sec.~\ref{sec:reconstruction}, except with small enough networks ($N=6$) to allow for exhaustive enumeration over all backbones $G^{(b)}\subseteq G$ to verify the optimality of Algorithm~\ref{alg:global} and Algorithm~\ref{alg:local}. Figure~\ref{fig:optimality} shows the compression ratio (Eq.~\ref{eq:compression-ratio}), obtained using both the greedy method (x-axis) and exact enumeration over possible backbones (y-axis), for all experimental trials shown in Fig.~\ref{fig:reconstruction}. We find that the greedy algorithm identifies an identical description length to exact enumeration in all cases up to machine precision.

\section{Additional Comparisons on Synthetic Networks}
\label{appendix:synthetic}
In this Appendix we provide additional tests to compare the different backboning methods with the synthetic network model of Sec.~\ref{sec:synthetic}. 

\begin{figure*}
    \centering
    \includegraphics[width=\textwidth]{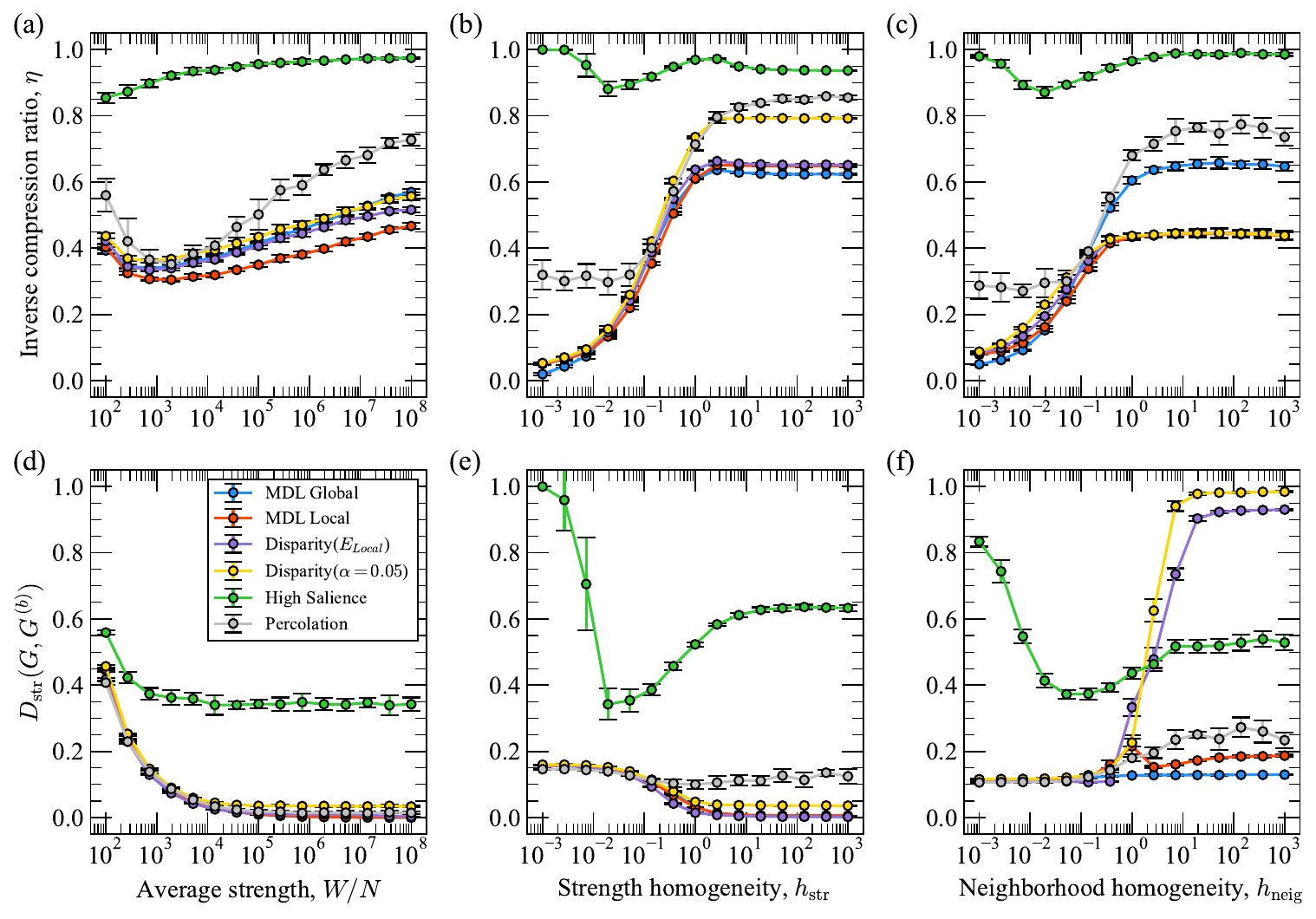}
    \caption{
    \textbf{Compression and node strength discrepancies among different backbones.}
    Top row: Inverse compression ratio (Eq.~\ref{eq:compression-ratio}) versus the \textbf{(a)} average node strength $W/N$, \textbf{(b)} level of homogeneity $h_{\text{str}}$ in the node strength distribution, and \textbf{(c)} level of homogeneity $h_{\text{neig}}$ in the weights within each node neighborhood. Bottom row: Panels \textbf{(d)}-\textbf{(f)} plot the Hellinger distance (Eq.~\ref{eq:hellinger}) between the original strength distribution and the backbone strength distribution against the same parameters. Markers indicate averages over 10 simulations from the synthetic network model described in Sec.~\ref{sec:synthetic}, and error bars represent one standard error in the mean.}
    \label{fig:synthetic-CRD}
\end{figure*}

In the top row of Fig.~\ref{fig:synthetic-CRD}, we can see that (as expected) the global MDL method is the most compressive among the global methods and the local MDL method is the most compressive among the local methods. Although the local methods tend to be more compressive in general, for high strength homogeneity (panel (b)) we find a slight compressibility advantage for the global method, suggesting that for networks with simple weight distributions one may favor the global MDL method over the local MDL method due to the additional model complexity of the latter. All methods exhibit worse compression as the weights become more homogeneous, since there is little statistical evidence for a separate backbone of high weight edges. In the bottom row we plot the Hellinger distance between the node strength distributions in the full network and the backbone (Eq.~\ref{eq:hellinger}), which shows similar performance for the global MDL method, local MDL method, and Disparity Filter with $E_{Local}$, while the Disparity Filter at $\alpha=0.05$ has slightly worse performance (e.g. a higher discrepancy in the strength distributions). In panel (f), however, we see that the MDL methods are much more robust to neighborhood weight heterogeneity fluctuations than the other methods, maintaining relatively similar strength distributions in the inferred backbones as the original networks. 

\begin{figure*}[h]
    \centering
    \includegraphics[width=0.9\textwidth]{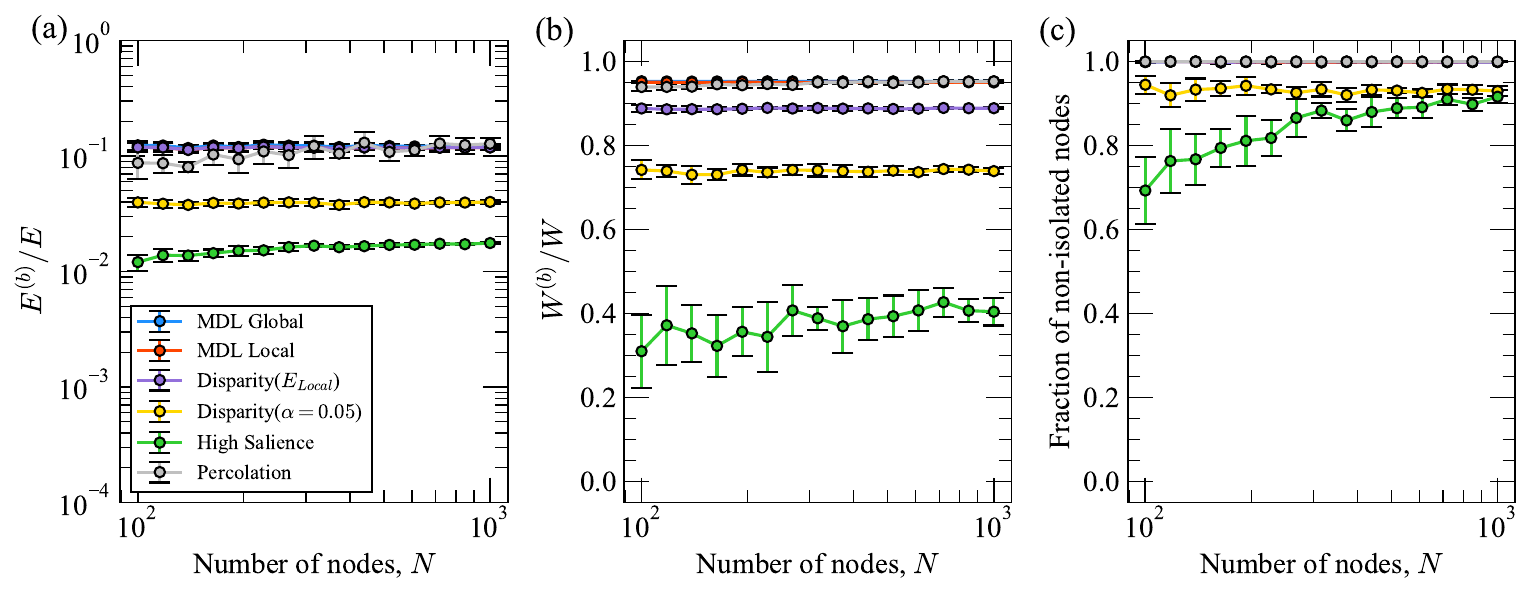}
    \caption{
    \textbf{Sizes of backbones for different network sizes.}
    \textbf{(a)} Fraction of edges retained in the backbone, \textbf{(b)} fraction of total weight retained in the backbone, and \textbf{(c)} fraction of non-isolated nodes in the backbone versus the number of nodes in the synthetic networks generated with the model in Sec.~\ref{sec:synthetic}. Markers indicate averages over 10 simulations from the synthetic network model described in Sec.~\ref{sec:synthetic}, and error bars represent one standard error in the mean.}
    \label{fig:synthetic-app-N}
\end{figure*}

\begin{figure*}[h]
    \centering
    \includegraphics[width=0.6\textwidth]{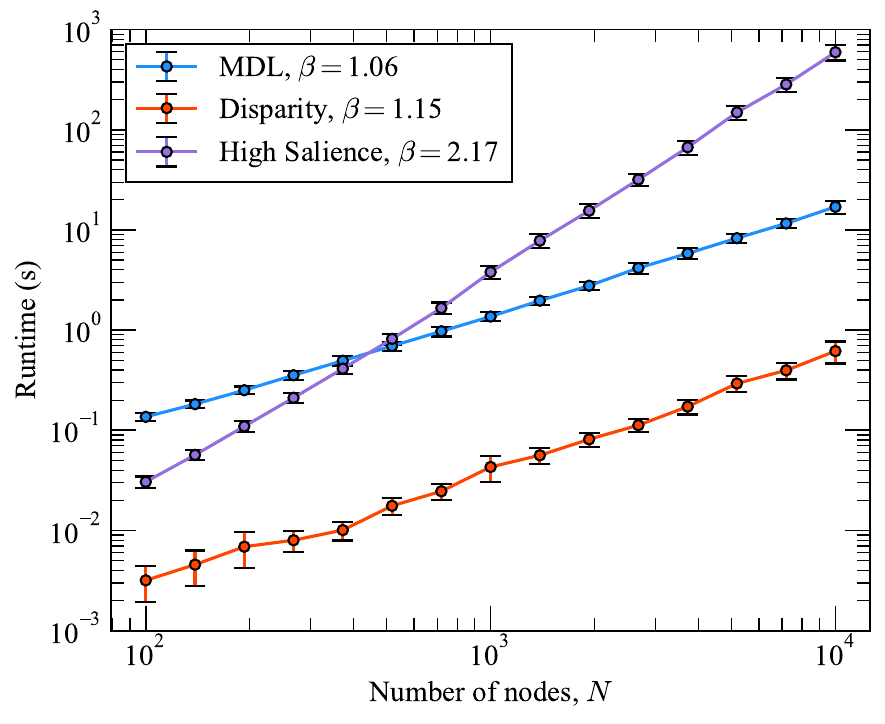}
    \caption{
    \textbf{Runtime scaling of different backboning methods.} Runtime per backbone evaluation (in seconds) versus the number of nodes $N$ for the synthetic network experiments in Sec.~\ref{sec:synthetic}. The global and local MDL backbones are computed together, as well as the Disparity Filter backbones. Best-fit slope values $\beta$ for linear fits of the form $\log (\text{Runtime})=\beta \log (N) + C$ are shown alongside each method in the legend. Markers indicate averages over 10 simulations from the synthetic network model described in Sec.~\ref{sec:synthetic}, and error bars represent one standard error in the mean.}
    \label{fig:synthetic-app-runtime}
\end{figure*}

Fig.~\ref{fig:synthetic-app-N} shows the sizes of the inferred backbones versus the number of nodes in the original network, suggesting a very weak relationship with this model parameter. And Fig.~\ref{fig:synthetic-app-runtime} plots the runtimes of a few methods for comparison, the slightly superlinear scaling of the MDL runtimes consistent with the $O(E\log E)$ scaling discussed in Sec.~\ref{sec:optimization}.

\section{Proof of Greedy Algorithm Optimality for Canonical Backbone Models}
\label{appendix:canonical-proof}

Here we prove the optimality of the greedy algorithm in Algorithm~\ref{alg:global} (and consequently Algorithm~\ref{alg:local}, when applied to individual neighborhoods) for a set of weight distributions in what is called the Natural Exponential Family (NEF) \cite{morris1982natural}, parametrized by
\begin{align}
P(w\vert \theta) = h(w)e^{\theta w - A(\theta)},    
\end{align}
where $h(w)$ is some function of only the weight and $A(\theta)=\log \int h(w)e^{\theta w}$ is the log normalizing constant (or log partition function). Due to the linearity in the exponent, in NEF distributions the edge weights in the backbone and non-backbone only interact with the model parameters $\theta$ through their sum $W^{(b)}$ in the model likelihood, allowing us to analyze the effect of changing $W^{(b)}$ with fixed $E^{(b)}$.

Using Eq.~\ref{eq:DLglobalfromBayes}, we have, up to constants independent of $W^{(b)}$,
\begin{align}
\mathcal{L}^{(\text{global})}_{C}
\propto  
- \log \left[\int P(\bm{\theta}_1) \prod_{e\in G^{(b)}}P(w_{e}\vert \bm{\theta}_1)d\bm{\theta}_1\int P(\bm{\theta}_0) \prod_{e\in G\setminus G^{(b)}} P(w_{e}\vert \bm{\theta}_0)d\bm{\theta}_0\right].
\end{align}
Using the arguments from Sec.~\ref{sec:optimization}, if we have 
\begin{align}
\Delta \mathcal{L}^{(\text{global})}_{C}(W^{(b)}\!\!\pluseq 1) &< 0~~\text{(integer weights)},\\
\pder{\mathcal{L}^{(\text{global})}_{C}}{W^{(b)}} &< 0~~\text{(continuous weights)},
\end{align}
given the greedy conditions 
\begin{align}
E^{(b)} &\leq \tilde E^{(b)},\\
\frac{W^{(b)}}{E^{(b)}} &\geq \frac{W}{E}\geq \frac{\tilde W^{(b)}}{\tilde E^{(b)}},    
\end{align}
then the greedy algorithms of Algorithm~\ref{alg:global} and Algorithm~\ref{alg:local} are optimal. We have used the notation $\tilde W^{(b)}=W-W^{(b)}$ and $\tilde E^{(b)}=E-E^{(b)}$ for brevity.

For the Poisson distribution with a maximum entropy exponential prior with scale parameter $\lambda$, we have that
\begin{align}
\mathcal{L}^{(\text{global})}_{C} \propto \log \frac{(E^{(b)}+\lambda)^{W^{(b)}+1}(\tilde E^{(b)}+\lambda)^{\tilde W^{(b)}+1}}{W^{(b)}!\tilde W^{(b)}!}.
\end{align}
Using the same argument as in Sec.~\ref{sec:optimization}, we have
\begin{align}
\Delta \mathcal{L}^{(\text{global})}_{C}(W^{(b)}\!\!\pluseq 1) 
=\log \frac{(E^{(b)}+\lambda)}{(W^{(b)}+1)}\frac{(\tilde W^{(b)}+1)}{(\tilde E^{(b)}+\lambda)}.
\end{align}
This is negative for $\frac{W^{(b)}+1}{E^{(b)}+\lambda}>\frac{\tilde W^{(b)}+1}{\tilde E^{(b)}+\lambda}$, which is asymptotically equivalent to the greedy condition $\frac{W^{(b)}}{E^{(b)}}>\frac{\tilde W^{(b)}}{\tilde E^{(b)}}$.

For the Geometric distribution, we showed in Appendix~\ref{appendix:equivalence} that the description length under a maximum entropy uniform prior is given by
\begin{align}
\mathcal{L}^{(\text{global})}_{C} = \mathcal{L}_M^{(\text{global})}(G^{(b)})
+ \log \frac{(W^{(b)}+1)(\tilde W^{(b)}+1)(W^{(b)})(\tilde W^{(b)})}{(W-E+1)(E^{(b)})(E-E^{(b)})}, 
\end{align}
which gives
\begin{align}
\Delta \mathcal{L}^{(\text{global})}_{C}(W^{(b)}\!\!\pluseq 1) = \log \frac{\tilde W^{(b)}-\tilde E^{(b)}}{W^{(b)}-E^{(b)}+1}\frac{W^{(b)}+2}{\tilde W^{(b)}+1},
\end{align}
where $\tilde E^{(b)}=E-E^{(b)}$ and $\tilde W^{(b)}=W-W^{(b)}$. This is negative for $\frac{W^{(b)}+2}{E^{(b)}+1}>\frac{\tilde W^{(b)}+1}{\tilde E^{(b)}+1}$, which is again asymptotically equivalent to the greedy condition $\frac{W^{(b)}}{E^{(b)}}>\frac{\tilde W^{(b)}}{\tilde E^{(b)}}$.

For the Exponential distribution with a maximum entropy exponential prior with scale parameter $\lambda$, we have that
\begin{align}
\mathcal{L}^{(\text{global})}_{C} \propto \log   \frac{(W^{(b)}+\lambda)^{E^{(b)}+1}(\tilde W^{(b)}+\lambda)^{\tilde E^{(b)}+1}}{E^{(b)}!\tilde E^{(b)}!}, 
\end{align}
which gives
\begin{align}
\pder{\mathcal{L}^{(\text{global})}_{C}}{W^{(b)}} = \frac{E^{(b)}+1}{W^{(b)}+\lambda} - \frac{\tilde E^{(b)}+1}{\tilde W^{(b)}+\lambda}. \end{align}
This is negative for $\frac{W^{(b)}+\lambda}{E^{(b)}+1}>\frac{\tilde W^{(b)}+\lambda}{\tilde E^{(b)}+1}$, which is asymptotically equivalent to the greedy condition $\frac{W^{(b)}}{E^{(b)}}>\frac{\tilde W^{(b)}}{\tilde E^{(b)}}$. 

We thus have that Algorithm~\ref{alg:global} and Algorithm~\ref{alg:local} are asymptotically optimal for the canonical backboning model with Poisson, Geometric, and Exponential distributions $P(w\vert \theta)$, archetypal examples of Natural Exponential Family (NEF) distributions.

\end{document}